\documentclass[useAMS,usenatbib,letterpaper]{mn2e}
\usepackage[totalwidth=480pt,totalheight=680pt]{geometry}
\usepackage{graphicx,amssymb}
\input psfig

\title[The Great Escape]
{The Great Escape: How Exoplanets and Smaller Bodies Desert Dying Stars}
\author[Veras, Wyatt, Mustill, Bonsor \& Eldridge]{Dimitri Veras$^{1}$\thanks{E-mail:
veras@ast.cam.ac.uk}, Mark C. Wyatt$^{1}$, Alexander J. Mustill$^{1}$, 
Amy Bonsor$^{1}$,
\newauthor and John J. Eldridge$^{1}$ \\
$^{1}$Institute of Astronomy, University of Cambridge, Madingley Road, Cambridge CB3 0HA}

\begin{document}

\date{Accepted 2011 July 6.  Received 2011 June 26; in original form 2011 May 29}

\pagerange{\pageref{firstpage}--\pageref{lastpage}} \pubyear{2011} 

\maketitle

\label{firstpage}

\begin{abstract}
Mounting discoveries of extrasolar planets orbiting post-main sequence
stars motivate studies aimed at understanding the fate of these
planets.  In the traditional ``adiabatic" approximation, a 
secondary's eccentricity remains constant during stellar mass loss.
Here, we remove this approximation, investigate the full two-body point-mass
problem with isotropic mass loss, and illustrate the resulting
dynamical evolution.  The magnitude and duration of a star's 
mass loss combined with a secondary's initial orbital characteristics 
might provoke ejection, modest eccentricity pumping,
or even circularisation of the orbit.  We conclude that Oort clouds and
wide-separation planets may be dynamically ejected from $1 M_{\odot}-7 M_{\odot}$ 
parent stars during AGB evolution.
The vast majority of planetary material which survives a
supernova from a $7 M_{\odot}-20 M_{\odot}$ progenitor will
be dynamically ejected from the system, placing limits
on the existence of first-generation pulsar planets.
Planets around $>20 M_{\odot}$ black hole progenitors may easily survive
or readily be ejected depending on the core collapse and superwind 
models applied.  
Material ejected during stellar 
evolution might contribute significantly to the free-floating planetary population.
\end{abstract}


\section{Introduction}

Understanding the formation and subsequent dynamical evolution of exoplanets 
has been a motivational hallmark for many observational and theoretical 
investigations.  However, extrasolar planets 
continue to be discovered in surprising and exotic environments,
and questions about the {\it endstate} of exoplanets are becoming increasingly relevant.
Few studies so far have modeled these systems, 
which often feature evolved and variable parent stars.
The rich dynamics therein fundamentally differ from
studies of planets around main sequence stars.

Examples of exoplanets which do not orbit main sequence stars 
are growing.  The first confirmed extrasolar planets were discovered around
a neutron star: specifically, the millisecond pulsar 
PSR1257+12 \citep{wolfra1992,wolszczan1994}.
The minimum masses of these three planets continue to be among
the lowest known to date, and two of these planets resonantly
interact.  \cite{sigurdsson2003} later discovered another pulsar
planet, around the binary radio millisecond pulsar PSR B1620-26.
Exoplanets are also thought to orbit white dwarfs and stars
with white dwarf companions.  In the first category, GD 66 
\citep{muletal2008,muletal2009}, GD 356 \citep{wicetal2010} and 
Gliese 3483 (Matt Burleigh, private communication)  
are planet-hosting stars.  In the second category, examples
are thought to include Gl 86 $=$ HD 13445 \citep{queetal2000,mugneu2005,lagetal2006},
HD 27442 \citep{butetal2001,chaetal2006}, and HD 147513 \citep{mayetal2004,desbar2007}.

Additionally, planets have been discovered orbiting stars
that have turned off of the main sequence but are not yet
stellar remnants.  \cite{siletal2007} discovered
a giant planet orbiting the extreme horizontal branch star V 391 
Pegasi, \cite{geietal2009} found a planet around the hot
subdwarf star HD 149382, \cite{leeetal2009} reported circumbinary planets
to the sdB+M eclipsing system HW Virginis, and \cite{setetal2010} 
suggested that the planet orbiting the red horizontal branch
star HIP 13044b might be of extragalactic origin.  Cataclysmic
variables are another class of systems which might harbor
planets, and recently, planets around the cataclysmic variables 
QS Vir \citep{qianetal2010a}, DP Leo \citep{qianetal2010b}
and HU Aqr \citep{qiaetal2011} have been reported.  

Prospects for discovering additional planets orbiting 
white dwarfs \citep{draetal2010,faeetal2011} and
extreme horizontal branch stars \citep{beasok2011} are promising,
and observational campaigns to do so have already been initiated
\citep{hogetal2009,benetal2010,schetal2010}.  The {\it Kepler} mission can 
detect even smaller bodies around white dwarfs \citep{disetal2010}.

Theoretical investigations regarding the evolution of planets
around post-main sequence stars have focused primarily on 
planet engulfment and interaction with the expanding stellar envelope,
both for exoplanets and specifically for the Earth.
\cite{villiv2007}, \cite{massarotti2008} and \cite{villiv2009} use particular
stellar evolutionary tracks to determine ranges of semimajor axes
at which planets are likely to be engulfed.
In this regime, tidal modelling has a significant effect on
the subsequent orbital evolution.  However, as summarized
by \cite{hansen2010}, the nature of tidal dissipation is poorly understood
and continues to yield different results depending on the model
and assumptions used.  For this reason, the fate of the Earth is uncertain.
\cite{sacetal1993}, \cite{rybden2001}, \cite{schcon2008} and \cite{iorio2010} 
all explore the fate of the Earth in light of the Sun's post main-sequence
mass loss, with differing results.  Alternatively, \cite{debsig2002} focus
on the stability of multi-planet systems and link stellar mass loss to
instability timescales. By doing so, they demonstrate how multiple 
planets beyond the reach of the star's expanding envelope might 
become unstable.

In this study, we consider just a single planet, or smaller body.
We perform a detailed analysis of the variable mass 
two-body problem and apply the results to a wide range of star-planet fates that
encompass all stellar masses $\lesssim 150 M_{\odot}$.  We focus 
on how stellar mass loss
affects the eccentricity of a planet or planetary material, a link
often ignored in previous studies.  As a result, we show that
planetary material can be ejected from a system based on mass loss
alone.  We then quantify for what combination of parameters we can
expect this behavior.

We start, in Section 2, by reviewing the history of the variable mass 
two-body problem and the corresponding equations of motion.  We then
analyze the orbital evolution in different mass loss regimes, 
determine where and when the traditionally-used adiabatic approximation holds,
and estimate when the planets would become unstable.  In Section 3, 
we apply the theory
to stars of all masses up to $150M_{\odot}$ in order to pinpoint realistic
systems which would yield instability.  We treat
five different mass regimes in separate subsections.  We then discuss the 
caveats, implications and potential extensions in Section 4, and 
conclude in Section 5.

\section{The General Two-Body Mass-Loss Problem}

\subsection{Overview}

Mass loss in the two-body problem, where both bodies are considered
to be point masses, has been
studied for over a century \citep[e.g.][]{gylden1884,mestschersky1893}.
This situation
is sometimes referred to as the ``Gyld\'{e}n-Mestschersky'' problem,
even though this particular case refers to both variable mass 
rates having the same functional dependence.
The more general problem takes many forms, or special cases, 
which are nicely outlined by Table 1 of \cite{razbitnaya1985}.  
One well-known form results from
the application of this general theory to binary 
stellar systems, a formalism pioneered by \cite{jeans1924}.  
The mass loss prescription which bears his name, 
$\dot{M} = -\kappa M^j$, where $M$ is mass and $\kappa$
and $j$ are constants, has been analytically and numerically
treated in many subsequent studies.  However, specific
applications of mass loss to planetary systems have received
little treatment.

Soon after the advent of computer-based numerical integrations, 
\cite{hadjidemetriou1963,hadjidemetriou1966a,hadjidemetriou1966b} revisited
and reformulated the problem in important ways.   \cite{hadjidemetriou1963} 
highlighted the subtlety with which mass loss must be treated
in order to retain physical interpretations of the evolution of orbital
elements.  He modeled mass loss as an additional acceleration that is a function of
a time- and mass-dependent velocity, and showed that for any isotropic mass loss prescription,
a planet's angular momentum $h$ satisfies:

\begin{equation}
h = {\rm constant} = \sqrt{G \mu  a \left(1 - e^2 \right)}
,
\label{conam}
\end{equation}

\noindent{where} $a$ refers to the semimajor axis, $e$ to the eccentricity,
and $\mu \equiv M_{\star} + M$.
The subscript ``$\star$'' refers to the star and the variables without
subscripts refer to the (lower-mass) secondary in the 
two-body system, which can be thought of as either a planet
or particle; we will use the term ``planet.''  Despite the conservation of angular momentum, no 
such conservation claim could be made about the total energy of the system.
\footnote{Because a system with isotropic mass loss will maintain its rotational symmetry, according
to Noether's Theorem, the angular momentum will be conserved.  Because the same system
does not exhibit time invariance, the energy of the system is not guaranteed
to be conserved.}
\cite{hadjidemetriou1966b} then significantly discovered that amidst
great mass loss, such as in a supernova, the eccentricity
of the secondary may increase, and eventually
lead to ejection from the system.  That finding is the foundation
for this work.  A subsequent series of papers \citep{verhulst1969,vereck1970,verhulst1972} 
provided an expansion of and comparison with Hadjidemetriou's results.
\cite{alcetal1986} then approached the ejection possibilities from a different perspective
by considering the effect of vigorous mass loss of white dwarf
progenitors on a comet.  Later, \cite{paralc1998} demonstrated how the 
asymmetric mass loss case yields a greater fraction of cometary ejections.

Despite the wide body of work on mass loss in the two-body problem\footnote{ 
\cite{rahomaetal2009} provides a detailed summary of additional results from
past papers, and \cite{plamuz1992} summarizes the ``use and abuse'' of using
a force to model mass loss.}, most studies continued to concentrate on binary stars.
\cite{debsig2002} helped break this trend by analyzing the planetary case
through the modelling of multiple planets orbiting a single star.  
They assumed the planets had
equal masses and initially circular orbits, and studied their motion
in the ``adiabatic'' approximation.  This approximation holds
when the {\it mass loss timescale is much greater than 
a planetary orbital timescale}.  In this approximation, 
the planet's eccentricity
is thought to remain nearly constant, and hence, from
Eq. (\ref{conam}),

\begin{equation}
\left(\frac{da}{dt}\right)_{\rm adiabatic} = 
- 
\frac{a}{\mu}\frac{d\mu}{dt}
.
\label{dadtgen2}
\end{equation}

However, in the general planetary case, the angular momentum is a function
of eccentricity, which is generally not constrained to be fixed.
Other complicating factors are: i) because planetary orbits which are changing
due to stellar mass loss are not closed, averaged orbital element expressions
can be misleading and counter-intuitive, although technically correct
\citep{iorio2010}, ii) in a single
phase of stellar evolution, mass loss is typically nonconstant 
(although monotonic) and may not be isotropic, iii) stellar 
mass evolution typically involves
multiple phases of mass loss on timescales which can vary by orders
of magnitude, and iv) several additional forces
due to stellar evolution, such as tides and dynamical friction
from the expanding envelope, might be necessary to model in order to 
describe the correct orbital evolution.

Here, we do not place restrictions a planet's semimajor axis, eccentricity or
orbital angles, but do take measures to focus our results.  We treat mass
loss as isotropic.  Tidal effects are unimportant in the regimes
we consider here, and so we can safely neglect those.  
To foster intuition for the mass loss problem,
and to obtain tractable results, our analytics assume
a constant mass loss rate throughout.
However, some of our analytical results are completely
independent of the mass loss rate assumed.  
The parameters for the example cases used in this Section 
were selected to best demonstrate different aspects of the 
motion of this general two-body problem with mass loss;
more realistic cases are presented in Section 3.  There,
we apply the theory presented 
here to just a single phase of stellar evolution, but do 
consider almost the entire phase space of
stellar mass. 

\subsection{Statement of Equations}

Although the equations of motion in terms of orbital elements
for the variable-mass two-body problem can be derived
from first principles, only a few authors
\citep[e.g.][]{hadjidemetriou1963,verhulst1969,deprit1983,li2008}
have stated them in full without averaging or approximation: 

\begin{eqnarray}
\frac{da}{dt} &=& 
- \frac{a \left(1 + e^2 + 2e \cos{f}\right)}
{1 - e^2} 
\frac{1}{\mu}\frac{d\mu}{dt}
\label{YESmoda}
     \\
\frac{de}{dt} &=& 
- \left(e + \cos{f} \right) 
\frac{1}{\mu}\frac{d\mu}{dt}
\label{YESmode}
    \\
\frac{di}{dt} &=& \frac{d\Omega}{dt} = 0   
     \\
\frac{d\omega}{dt} &=&  \frac{d\varpi}{dt} =   
-\frac{\sin{f}}{e}
\frac{1}{\mu}\frac{d\mu}{dt}
\label{YESmodom}
        \\
\frac{df}{dt}   &=& -\frac{d\varpi}{dt} +
\frac{ n
\left( 1 + e \cos{f} \right)^2 }
{\left(1 - e^2 \right)^{3/2}}
\label{YESmodf}
\end{eqnarray}

\noindent{where} $i$ is the inclination, $\Omega$ is the 
longitude of ascending node, 
$\varpi$ is the longitude of pericenter, $\omega$ is the argument of 
pericenter and $f$ is the true anomaly.  Equations (\ref{conam}), (\ref{YESmoda})
and (\ref{YESmode}) are self-consistent and may be derived from
one another with help from the vis-viva equation.  The time
derivative of position in terms of orbital elements
and the statement of the conservation of angular momentum in 
polar coordinates give Eqs. (\ref{YESmodom}) and (\ref{YESmodf}).

These equations may also be derived from more general considerations.
Gauge theory is a basis from which one may obtain sets of equations
such as Lagrange's planetary equations and Gauss' Planetary Equations
by defining just a single perturbative acceleration to the
classic two-body problem, and a gauge velocity.
The formulation of the
theory with regard to planetary dynamics as well as extensive 
descriptions can be found in 
\cite{efrgol2003,efrgol2004}, \cite{gurfil2004}, \cite{efroimsky2005a},
\cite{efroimsky2005b,efroimsky2006}, \cite{gurfil2007}
and \cite{gurbel2008}.  
\cite{hadjidemetriou1963} showed that the sum of the 
isotropic mass variation of 
both bodies is equivalent to a perturbative force with an 
acceleration of $\Delta\vec{A} = -(1/2)(d\mu/dt)(1/\mu) \vec{v}$,
where $\vec{v}$ is velocity.
This acceleration yields Eqs. (\ref{YESmoda})-(\ref{YESmodf}) 
directly for a zero gauge.  

Every variable 
in Eqs. (\ref{YESmoda})-(\ref{YESmodf}) is considered to be
a function of time.  The mean motion, $n$, is equal to 
$G^{1/2} \mu^{1/2}/a^{3/2}$, where $G$ is treated as 
the standard gravitational constant.  
Although we use $\mu$
throughout this work to emphasize how the 
motion is affected by the sum of the mass loss
(or gained) by both bodies, the value
of the planetary mass and how it changes with
time has a negligible effect on the results for $M_{\star} \gg M$.
For a $1 M_{\odot}$ star, if one assumes a planetary
mass of $\sim$10 Jupiter masses, which is on the order of the theoretical
upper bound, then $M/M_{\star} \sim 1\%$.  

The planet's true longitude, $\theta$, varies according to:

\begin{equation}
\frac{d\theta}{dt}   = 
\frac{n
\left( 1 + e \cos{f} \right)^2 }
{\left(1 - e^2 \right)^{3/2}}
,
\end{equation}

\noindent{which} is not explicitly dependent on the
mass loss rate and hence is equivalent to the case of
no mass loss.  This equation demonstrates that from the point
of view from a fixed reference direction, the secondary
will continue to circulate around a star that is losing
mass as long as the secondary remains bound. 

For completeness, 
we consider the evolution of
other traditionally-used orbital parameters.
The planet's eccentric anomaly, $E$, will vary according to:

\begin{equation}
\frac{dE}{dt} = \frac{n\left(1 + e \cos{f}\right)}{1-e^2}
+
\frac{\sin{f}}{e\sqrt{1-e^2}}
\frac{1}{\mu}\frac{d\mu}{dt}
.
\label{EccAna}
\end{equation}

\noindent{Note} that the right-hand sides of 
Eqs. (\ref{YESmoda})-(\ref{EccAna}) 
may be expressed in terms of the eccentric anomaly
instead of the true anomaly.  The planet's mean motion 
will vary according to:

\begin{equation}
\frac{dn}{dt} = \frac{n\left(2+e^2 + 3e \cos{f}\right)}{1-e^2}
\frac{1}{\mu}\frac{d\mu}{dt}.
\label{meanmot}
\end{equation}

\noindent{The} planet's mean anomaly, $\Pi$, can be expressed as an explicit
function of time by use of the ``time of pericenter'', $\tau$:

\begin{equation}
\frac{d\Pi}{dt} = n + n\left(t - \tau \right)\frac{\left(2+e^2 + 3e \cos{f}\right)}{1-e^2}
\frac{1}{\mu}\frac{d\mu}{dt}
\end{equation}

\noindent{or}, through Kepler's Equation, as:

\begin{equation}
\frac{d\Pi}{dt} = n + \frac{\sqrt{1-e^2} \sin{f} \left(1 + e^2 + e \cos{f}\right)}{e \left(1 + e \cos{f} \right)} 
\frac{1}{\mu}\frac{d\mu}{dt}
,
\end{equation}

\noindent{which} is explicitly independent of time.  Finally, the mean 
longitude, $\lambda$, changes with time according to 

\begin{eqnarray}
\frac{d\lambda}{dt} &=& \frac{d\Pi}{dt} + \frac{d\varpi}{dt} 
= 
\nonumber
\\
&&n - \frac{d\varpi}{dt} \left[ \frac{\sqrt{1-e^2}\left(1 + e^2 + e \cos{f}\right)}{1+e\cos{f}} - 1\right]
.
\end{eqnarray}

\noindent{Throughout} this paper, we denote initial values with
the subscript ``0''.

\subsection{Parametrizing Mass Loss}

Suppose the mass loss 
rate is constant and equal to $-\alpha$,
such that $\alpha > 0$.
Then $\mu = G \left( \mu_0 - \alpha t \right)$,
and

\begin{equation}
\left(\frac{1}{\mu}
\frac{d\mu}{dt}
\right)
= -\left( \frac{\mu_0}{\alpha} - t \right)^{-1}
.
\end{equation}

We can better quantify adiabaticity and various
regimes of motion due to mass loss by
defining a dimensionless ``mass loss index'', $\Psi$:

\begin{eqnarray}
\Psi &\equiv& \frac{\alpha}{n \mu}
\nonumber
\\
&=& \frac{1}{2\pi} 
\left( \frac{\alpha}{1 M_{\odot}/{\rm yr}}\right)
\left( \frac{a}{1 {\rm AU}}\right)^{\frac{3}{2}}
\left( \frac{\mu}{1 M_{\odot}}\right)^{-\frac{3}{2}}
.
\label{mlindex}
\end{eqnarray}

This parameter provides a scaled ratio of the orbital
period to the mass loss timescale.  The initial value of the index
as $\Psi_0$.  Hence, the time evolution of $\Psi$ is governed
by:

\begin{equation}
\frac{d\Psi}{dt}
=
-3 \Psi 
\left( \frac{1+e \cos{f}}{1-e^2} \right)
\frac{1}{\mu}
\frac{d\mu}{dt}
.
\end{equation}

\noindent{}When $\Psi \ll 1$, a system can be considered ``adiabatic'',
the case we treat first.

\subsection{``Adiabatic'' Regime Evolution}

\subsubsection{Adiabatic Eccentricity Evolution}

We begin analyzing the equations of motion by 
first considering Eq. (\ref{YESmode}),
because all nonzero equations of motion
explicitly include $e$ in some manner. Note
importantly that the equation 
demonstrates that an initially circular planet will not remain
on a circular orbit, and that the planet's eccentricity 
will undergo oscillations on
orbital timescales when the parent star loses mass.

\begin{figure}
\centerline{
\psfig{figure=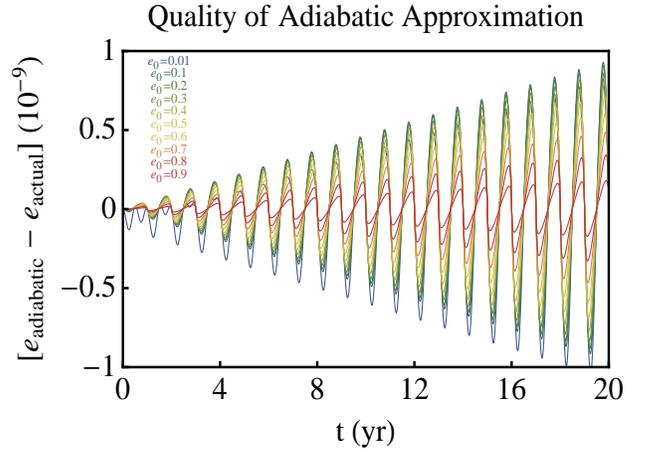,height=6cm,width=8.5cm} 
}
\caption{Analytic approximation to the eccentricity 
evolution in the adiabatic regime. 
Shown here is the difference in eccentricity of a planet
evolving according to Eq. (\ref{adiab}) compared to
Eqs. (\ref{YESmoda})-(\ref{YESmodf}).  The planet
is located at $a_0=1$ AU from a $\mu_0 = 1 M_{\odot}$ star
which is losing mass at the rate of
a rate of $\alpha = 10^{-5} M_{\odot}$/yr ($\Psi_0 \approx 1.6 \times 10^{-6}$).
The differently coloured lines from the top of
each crest moving downward correspond to
$e_0 = 0.01,0.1,0.2,0.3,0.4,0.5,0.6,0.7,0.8,$ and $0.9$
respectively.
}
\label{smallevar}
\end{figure}

We can solve Eq. (\ref{YESmode}) by noting that
in the adiabatic approximation ($\Psi \ll 1$),
the first term
in Eq. (\ref{YESmodf}) is considered
to be negligible compared to the
second term ($=d\theta/dt$), because the first
term is proportional to the mass loss rate.
Further, $\mu$ is assumed to remain fixed 
over the course of one orbit.  Hence, in this
regime, Eq. (\ref{YESmode})
may be integrated directly over the true anomaly, 
with the result:

\begin{equation}
e_{\rm adiabatic} = e_0 + 
\Psi_0
\frac{ \left( 1-e_{0}^2 \right)^{\frac{3}{2}} \sin{f}}{1 - e_0 \cos{f}}
.
\label{adiab}
\end{equation}

According to Eq. (\ref{adiab}), after each orbit
the eccentricity will return to its initial value.
During the orbit, the amplitude of $\left(e_{\rm adiabatic} - e_0\right)$ is
$\Psi_0 (1 - e_{0}^2) \propto \alpha$.  
Thus, assuming a current value of 
$\alpha_{\odot} \approx 10^{-13}$/yr,
the Earth's eccentricity is raised by about $10^{-14}$  
each year due to the Sun's mass loss.

\begin{figure*}
\centerline{
\psfig{figure=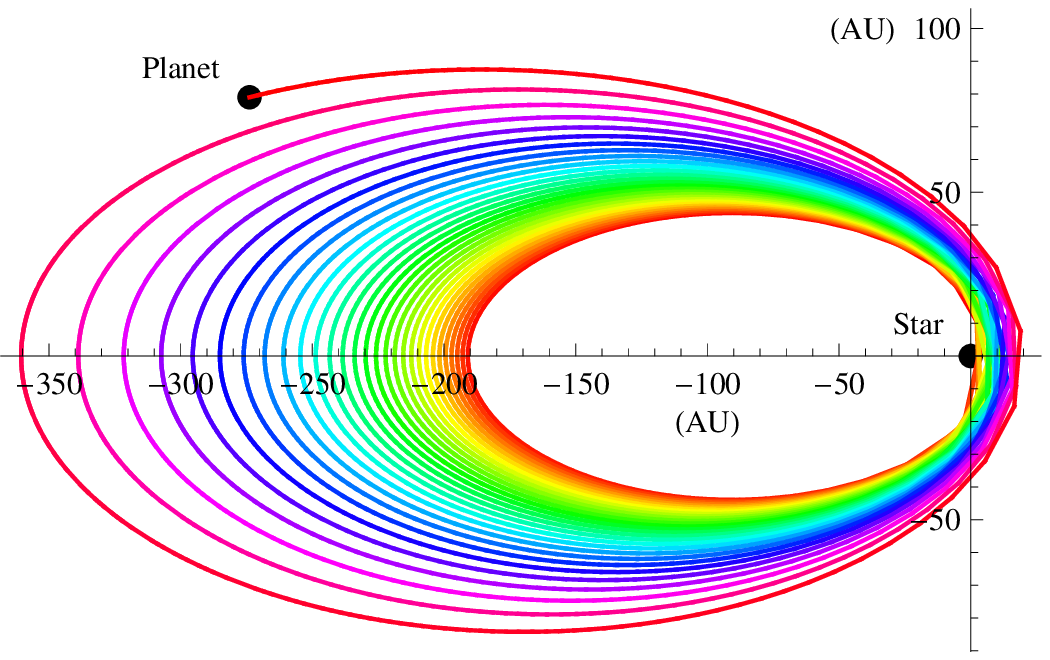,height=7cm,width=8.5cm}
\psfig{figure=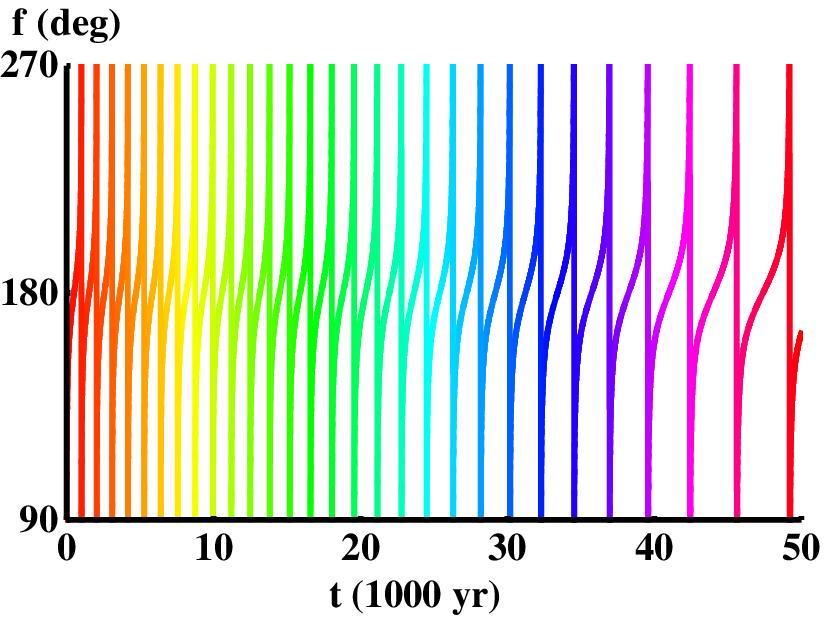,height=7cm,width=8.5cm}
}
\caption{
The adiabatic regime.  The position in space  
({\it left panel}) and the
evolution of the true anomaly ({\it right panel})
of a planet (or belt particle) that is being 
pushed outward due to stellar mass loss.
The colours on the curves indicate evolution at the same
points in time, and the vertical lines of true anomaly
indicate circulations of the angle.
The star has an initial mass of 
$\mu_0 = M_{\star} = 1 M_{\odot}$ and is losing mass
at the rate of $\alpha = 1 \times 10^{-5} M_{\odot}$/yr.
The planet begins on a highly eccentric orbit ($e_0=0.9$)
at $a_0=100$ AU, with $f_0=0^{\circ}$.
Hence, $\Psi_0 = 0.0016$.
Notice that as the planet moves outward and its
mean motion decreases, the circulation period
of the true anomaly decreases as well.
}
\label{beforebif}
\end{figure*}

Fig. \ref{smallevar} demonstrates the accuracy
of Eq. (\ref{adiab}) when compared with the evolution
from the full equations of motion 
(Eqs. \ref{YESmoda}-\ref{YESmodf}) for a $a_o=1$ AU
planet orbiting a $\mu_0 = 1 M_{\odot}$ star
which is losing mass at the rate of
$10^{-5} M_{\odot}$/yr (8 orders of magnitude greater
than $\alpha_{\odot}$).  The agreement is excellent
over the course of a single orbit. Over time, 
the approximation gradually worsens, as the evolution
of $\Psi_0$ is not taken into account in Eq. (\ref{adiab}).

\subsubsection{Adiabatic Semimajor Axis Evolution}

We now consider
the semimajor axis evolution from Eq. (\ref{YESmoda}).  
Note from the equation that for any mass loss, the semimajor
axis can never decrease.  

In the adiabatic regime, 
the semimajor axis
is traditionally evolved according to Eq. (\ref{dadtgen2}).
Note, however, that the equation does not
follow from Eq. (\ref{YESmoda}) if $e \ne 0$.
Yet, when the semimajor axis is
averaged over one orbital period, the eccentricity
terms vanish and Eq. (\ref{dadtgen2}) is recovered.
The solution of this equation is:

\begin{equation}
a_{\rm adiabatic} =  a_0 
\left( 1 - \frac{\alpha t}{\mu_0}  \right)^{-1}
.
\label{secondasol}
\end{equation}

Therefore, an adiabatically evolving planet will, for example, 
double its orbital separation if its parent $1 M_{\odot}$ star
constantly loses mass at the rate of 
$\alpha = 5 \times 10^{-9} M_{\odot}$/yr over 100 Myr.
In a different example, a $2 M_{\odot}$ star is expected 
to lose at most $\approx 70\%$
of its initial mass.  Therefore, {\it if} all this mass is 
lost adiabatically, then orbiting planets can expect to 
increase their semimajor axis by at most a factor of $\approx 3.3$.

\subsubsection{Adiabatic Orbital Angle Evolution}

Turning to other orbital parameters, the longitude
of pericenter is a typically secular
feature of multi-planet extrasolar systems.  Its 
variational timescale is often on 
the order of thousands of orbits.
During stellar evolution, however,
Eq. (\ref{YESmodom}) demonstrates that the variation
in a planet's longitude of pericenter 
is quick (on orbital timescales), and changes sign 
over each orbital period.
To be consistent with the adiabatic approximation,
in which $d\omega/dt \approx 0$ in Eq. (\ref{YESmodf}),
then

\begin{equation}
\varpi_{\rm adiabatic} = \varpi_0
.
\end{equation}

Because $d \varpi/dt$ is assumed to be zero, 
the value of $f_{\rm adiabatic}$ follows the same evolution
as $f$ would in the two-body problem with no mass
loss.

We can obtain an adiabat for $n$ from 
Eq. (\ref{meanmot}) under the same
assumptions that were used to derive 
Eqs. (\ref{dadtgen2}) and (\ref{secondasol}):

\begin{equation}
n_{\rm adiabatic} = n_0 
\left(
1-
\frac{\alpha t}{\mu_0}
\right)^2
.
\label{finaln}
\end{equation}

\noindent{Thus,} in the adiabatic approximation, the mean motion is 
a monotonically decreasing function.  In the same
example system from Section 2.4.2, with $\mu_0 = 1 M_{\odot}$,
and $\alpha = 5 \times 10^{-9} M_{\odot}$/yr, after $t = 100$ Myr the planet's
mean motion would decrease by a factor of 4.  This result is
expected from Kepler's 3rd law with a halved stellar mass a
doubled semimajor axis.  For $2 M_{\odot}$ stars, the final Keplerian
period of a planet when mass loss has ceased would be enhanced
from its initial period by a factor of at most $\approx 11$.

\subsubsection{Adiabatic Evolution in Space}

In space, adiabatic evolution corresponds to a planet orbiting
in an outward spiral pattern.  Figure \ref{beforebif} displays
such an orbit (for $\Psi_0 \approx 0.0016$), which is not closed.  
After each cycle of true longitude, the eccentricity 
does return to its initial osculating value.
The semimajor axis is seen to increase by as
much as $10\%$ of $a_0$ per orbit.
The increase
in orbital period can be 
linked with the circulation timescale of $f$.

A highly eccentric planet might make close passes
to the star, close enough to be affected by tides
and the stellar envelope.  In order
to determine if the planet is more or less likely
to suffer these encounters from mass loss,
consider the evolution of the pericenter, $q$, 
of the planet:

\begin{equation}
\frac{dq}{dt} = 
-\frac{a \left(1-e\right) \left(1-\cos{f}\right)}{1+e} 
\frac{1}{\mu}\frac{d\mu}{dt}
.
\label{qevol}
\end{equation}

\noindent{Equation} (\ref{qevol}) demonstrates that 
the pericenter {\it monotonically increases with
stellar mass loss}.  The left panel of 
Fig. \ref{beforebif} corroborates
this relation.  Therefore, if a planet ``outruns'' the 
star's expanding envelope, then one can neglect the
envelope's influence on the planet.  

\subsection{Regime Transition}
\subsubsection{The Breaking of Adiabaticity}

Equation (\ref{finaln}) has important implications for the 
dynamical system, as 
mean motion is inversely proportional to the Keplerian period.
Hence, as a star loses mass, and pushes a planet radially outward,
the mean motion decreases, and {\it eventually the orbital period
will be comparable to the mass loss timescale} ($\Psi \sim 1$).  
More precisely,
$d\theta/dt$ (which is proportional to the mean motion) will 
eventually become equal to $(-d\varpi/dt)$ (which is proportional
to the mass loss timescale).  At this bifurcation point in the
dynamics, Eq. (\ref{YESmodf}) 
demonstrates that the true anomaly becomes momentarily stationary. 
At this point, one can claim that adiabaticity is broken. 

\begin{figure*}
\centerline{
\psfig{figure=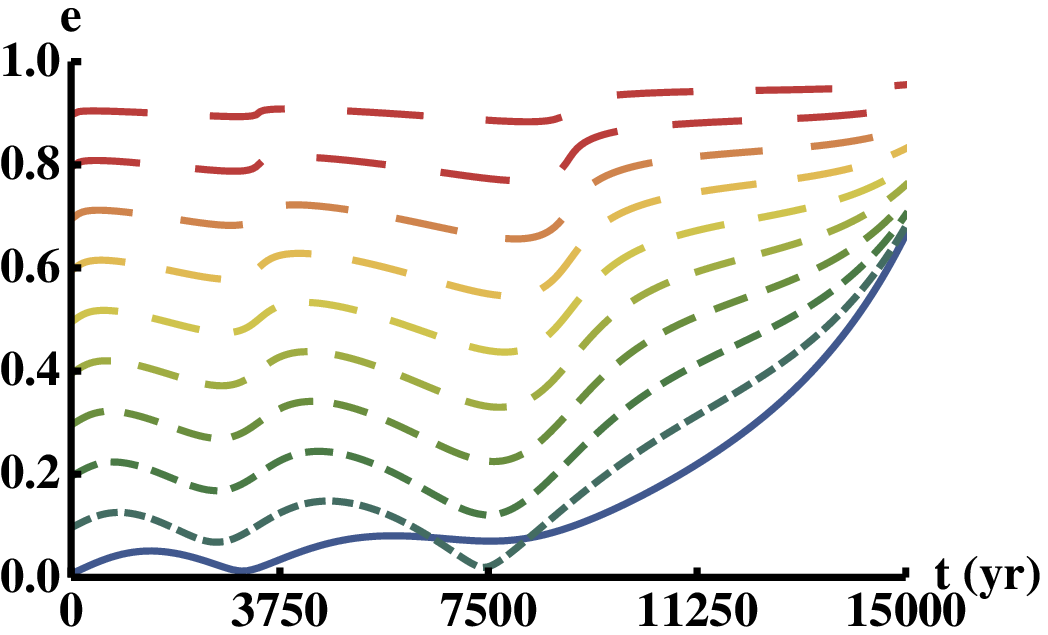,height=6cm,width=8.5cm} 
\psfig{figure=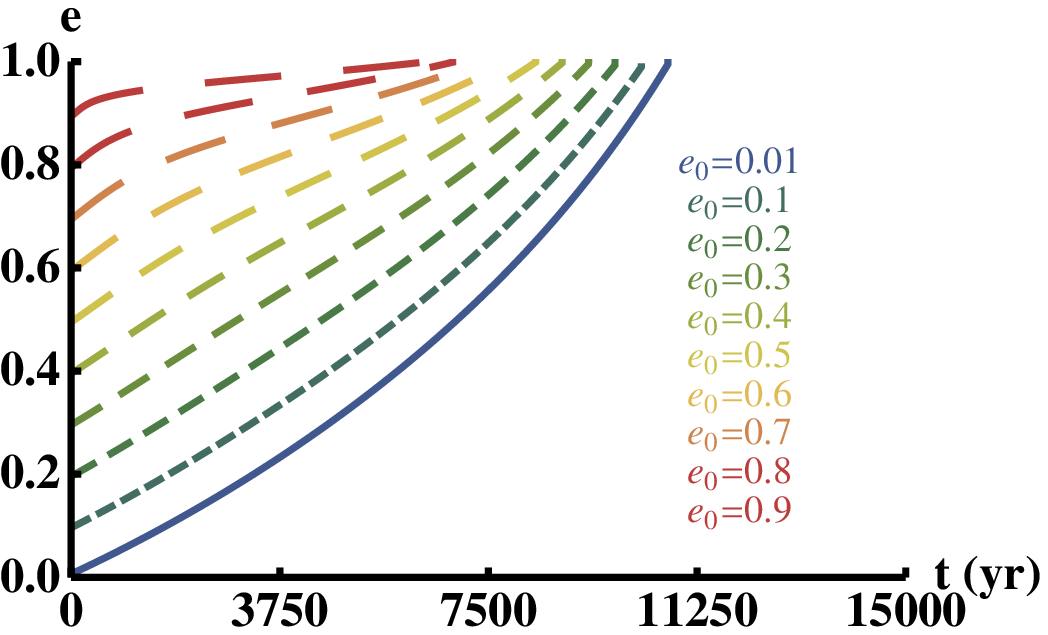,height=6cm,width=8.5cm}
}
\caption{The breaking of adiabaticity.
Shown is the eccentricity evolution over $10^4$ years 
of evolution of
$a_0 = 200$ AU planets (or belt particles; {\it left panel}) 
and 
$a_0 = 10^3$ AU planets (or belt particles; {\it right panel})
orbiting a $\mu_0 = M_{\star} = 1 M_{\odot}$ star losing mass at
a rate of $\alpha = 5 \times 10^{-5} M_{\odot}$/yr. 
The initial mass loss index for the systems in the left and
right panels are respectively $\Psi_0 \approx 0.023$ and
$\Psi_0 \approx 0.25$, values which are close to the transition
point in the dynamics between the adiabatic and runaway
regimes.  Here, $f_0=0^{\circ}$.
The lines with an increasing
dash length represent $e_0$ values of
$0.01,0.1,0.2,0.3,0.4,0.5,0.6,0.7,0.8,$ and $0.9$, respectively.
In the right panel, all planets are ejected from
the system within $10^4$ yr except the
two planets with the lowest initial eccentricity.
}
\label{largeevar}
\end{figure*}

\begin{figure*}
\centerline{
\psfig{figure=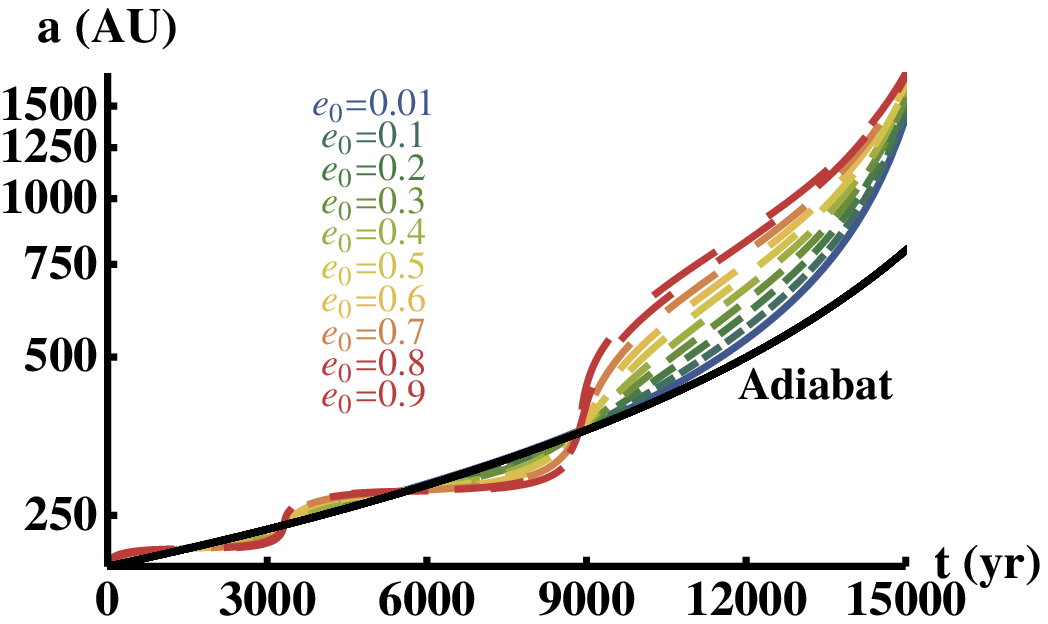,height=6cm,width=8.5cm} 
\psfig{figure=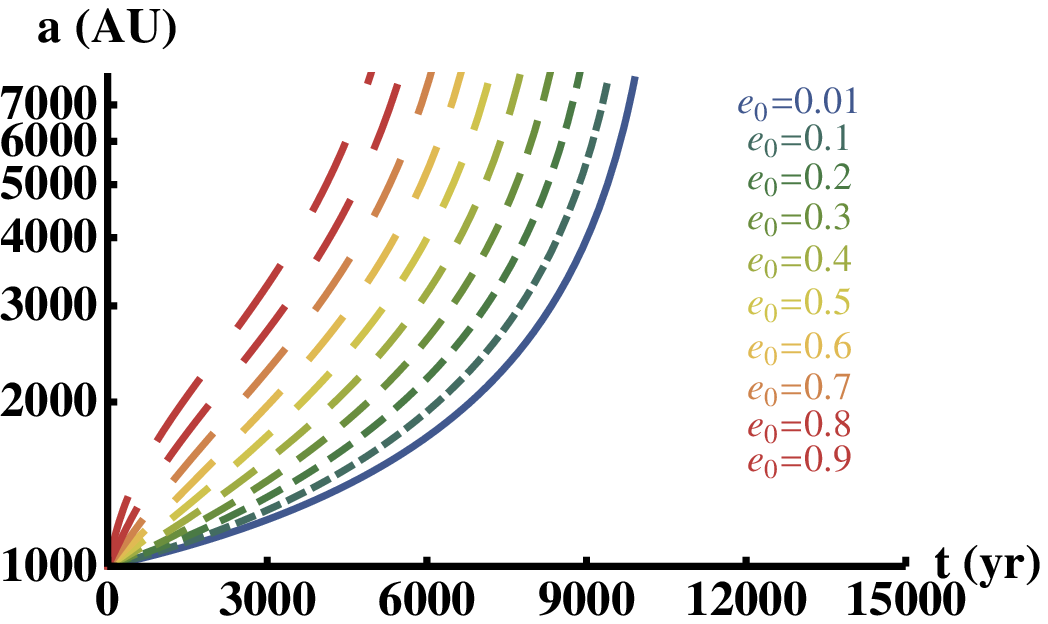,height=6cm,width=8.5cm}
}
\caption{The breaking of adiabaticity for the same
two systems in Fig. \ref{largeevar}.  In the left panel,
note how the eccentric planets oscillate about the
adiabat until reaching the runaway regime.
In the right panel, the planets begin at $t=0$
just off of the adiabat, and quickly settle
into the runaway regime.
}
\label{avar}
\end{figure*}

Note that in the adiabatic regime, $f$ circulates.
At and beyond the bifurcation point, $df/dt$ 
instead begins to librate.
The effect of a librating $f$ on $da/dt$ and 
$de/dt$ is pronounced, quick and runaway.
The eccentricity and semimajor axes evolution 
undergo a qualitative change, and the rate of increase
in the latter is pronounced. Therefore, we 
denote this regime as the ``runaway'' regime.
We wish to investigate this regime transition,
and do so first qualitatively through 
Figs. \ref{largeevar} and \ref{avar}.
These figures model a $1 M_{\odot}$ star which is
losing mass at a relatively high rate of 
$\alpha = 5 \times 10^{-5} M_{\odot}$/yr over $1.5 \times 10^4$ yr.
After this amount of time, the star will have lost $75\%$
of its initial mass.  These values are chosen for demonstration
purposes, as typical $1 M_{\odot}$ stars will lose $\approx 35\% - 62\%$
of their mass en route to becoming a white dwarf.  We model
more realistic systems in Section 3.

Figure \ref{largeevar} 
illustrates the approach to and onset of
adiabaticity breaking, with $\Psi_0 = 0.023$ (left panel)
and $\Psi_0 = 0.25$ (right panel).  In the first case, the
eccentricity ceases to remain approximately constant
and can start to oscillate on the order of a tenth. 
In the second case, where $\Psi$ quickly assumes values on
the order of unity, planets evolve in the
runaway regime and may achieve hyperbolic 
orbits.

Figure \ref{avar} showcases
the semimajor axis evolution for the
same systems in Fig. \ref{largeevar}.
In the left panel of Fig. \ref{avar},
the curves of initial eccentricity
break away from the adiabat, increasing
at a steeper rate than the adiabat.
In the right panel, at $t=0$, the
systems are just beyond the adiabat
and are ``running away'' from the star.
For a constant mass loss
that turns on and off nearly instantaneously
(such as in a supernova),
planets might not ever evolve
adiabatically, and begin
their life in the runaway regime.
Note that unlike eccentricity,
the semimajor axis is always increasing,
even when oscillating about the adiabat.
In the runaway regime, the departure
from the adiabat is drastic; 
the right panel shows that the planet
may increase its semimajor axis 
by many factors before
achieving a hyperbolic orbit.

The resulting increase in $a$ will cause the mean motion
term in Eq. (\ref{YESmodf}) to decrease further.
In the limiting case where $d\omega/dt \gg n$, 
the libration amplitude will approach zero, and
the true anomaly will become nearly stationary.  As a result,
the $\cos{f}$ term in Eq. (\ref{YESmode}) becomes constant,
and $de/dt$ becomes linear in $e$, causing a positive 
feedback effect that is characteristic of the runaway
regime.

\begin{figure*}
\centerline{
\psfig{figure=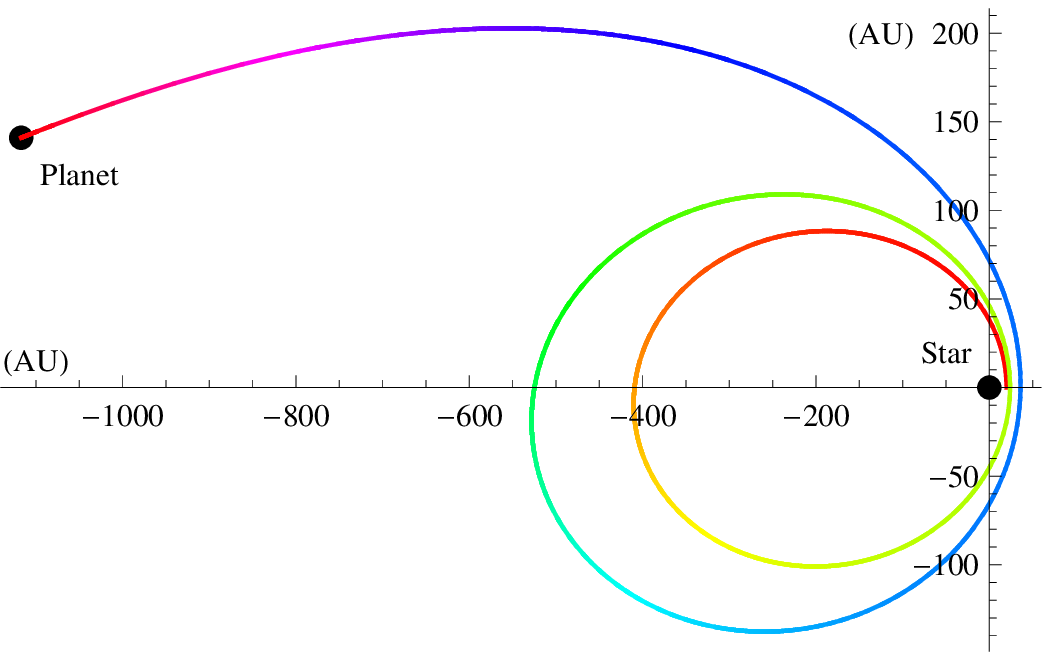,height=6cm,width=8.5cm} 
\psfig{figure=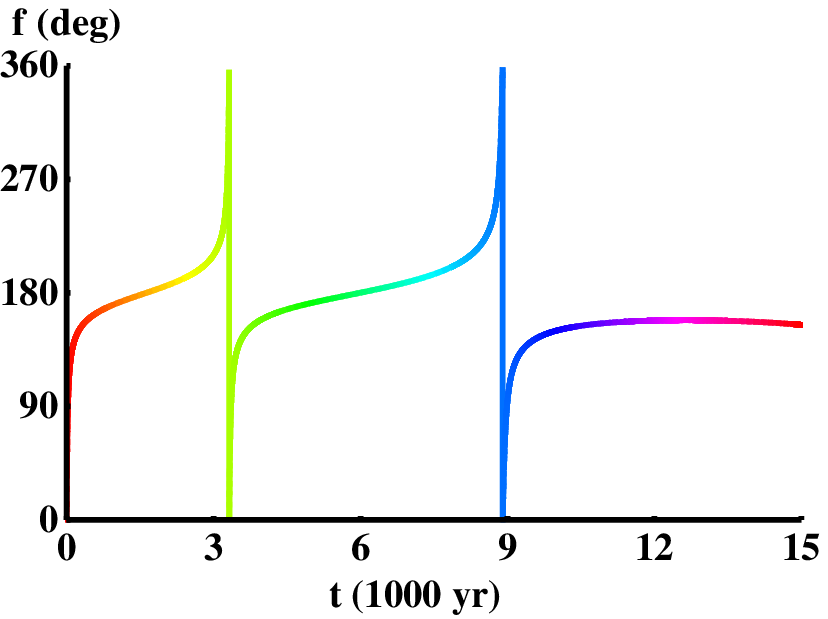,height=6cm,width=8.5cm}
}
\caption{
The position in space 
({\it left panel}) and the
evolution of the true anomaly ({\it right panel})
of the $e_0=0.9$ planet from the system
in the left panels of Figs. \ref{largeevar} and \ref{avar}.
At $t \approx 1.2 \times 10^4$ yr, the planet stops
circulating and starts to head out of the system
as the true anomaly becomes stationary.
}
\label{xynext}
\end{figure*}

An orbit that is transitioning out of adiabaticity
will not complete its final orbit around the star, as the 
true anomaly is no longer circulating.  
Figure \ref{xynext} illustrates the resulting motion in 
space.  The system selected is the highest eccentricity
($e_0 = 0.9$) curve from the left panels of 
Figs. \ref{largeevar} and \ref{avar} ($\Psi_0 = 0.023$).  The system
reaches the bifurcation point at $\approx 1.2 \times 10^4$ yr,
as one can read off from the right panel.  Now we explore
this bifurcation point analytically.

\subsubsection{Characterising the Bifurcation Point}

The bifurcation point, as we defined in the last subsection, is the first moment when $d\theta/dt = -d\varpi/dt$.  At this moment, 
from Eq. (\ref{YESmodf}), $df/dt=0$, and

\begin{equation}
\Psi_{\rm bif} = \frac{e_{\rm bif} \left( 1+ e_{\rm bif} \cos{f_{\rm bif}} \right)^2}
{\sin{f_{\rm bif}} \left( 1-e_{\rm bif}^2 \right)^{3/2}  }
.
\label{psibif}
\end{equation}

For the majority of possible values of eccentricity and true anomaly, $\Psi_{\rm bif} \approx 0.1-1.0$.  There are an infinite number of triples $(\Psi_{\rm bif}, e_{\rm bif}, f_{\rm bif})$ that satisfy Eq. (\ref{psibif}).  We cannot determine any of these three values from the initial conditions, although one may approximate $e_{\rm bif} \approx e_0$ from the adiabatic approximation.  However, at this point in the planet's evolution, $e_{\rm bif}$ might differ by over 0.1 from $e_0$.

The functional form of Eq. (\ref{psibif}) suggests that for a given $\Psi_{\rm bif}$ and a given $e_{\rm bif}$, there might be more than one value of $f_{\rm bif}$ which satisfies the equation.  We now investigate this possible multiplicity further by considering the extremities of $\Psi_{\rm bif}$ with respect to $e_{\rm bif}$ and $f_{\rm bif}$.  There are six values of $f_{\rm bif}$ which satisfy $d\Psi_{\rm bif}/df_{\rm bif} = 0$, five of which are unphysical.  The one physical solution is:

\begin{equation}
f_{\rm bif,min} = \cos^{-1}{\left[ \frac{1-\sqrt{1+8 e_{\rm bif}^2}}
{2e_{\rm bif}} \right]},
\label{fcrit}
\end{equation}

\noindent{where} $90^{\circ} \le f_{\rm bif} \le 270^{\circ}$.   Let the value of $\Psi_{\rm bif}$ at $f_{\rm bif} = f_{\rm bif,min}$ be denoted as $\Psi_{\rm bif,fmin}$.  Then, for a given $\Psi_{\rm bif}$ and a given $e_{\rm bif}$, the number of values of $f_{\rm bif}$ which satisfy Eq. (\ref{psibif}) are:

\begin{eqnarray}
0& \ \ {\rm values} \ \ {\rm of}  \ \ f_{\rm bif} \ \ {\rm if}& \ \ \Psi_{\rm bif} < \Psi_{\rm bif,fmin}
\label{firb}
\\
1& \ \ {\rm value} \ \ {\rm of} \ \ f_{\rm bif} \ \ {\rm if}& \ \ \Psi_{\rm bif} = \Psi_{\rm bif,fmin}
\\
2& \ \ {\rm values} \ \ {\rm of} \ \ f_{\rm bif} \ \ {\rm if}& \ \ \Psi_{\rm bif} > \Psi_{\rm bif,fmin}
.
\label{lasb}
\end{eqnarray}

\begin{figure}
\centerline{
\psfig{figure=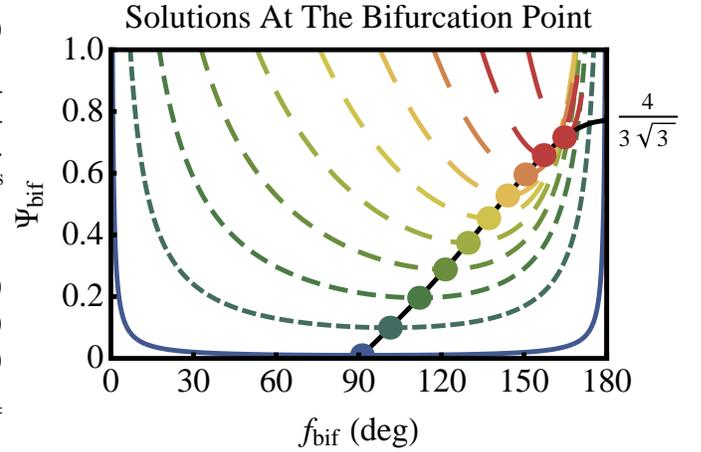,height=6cm,width=9.5cm}
}
\caption{
Values of $(\Psi_{\rm bif}, e_{\rm bif}, f_{\rm bif})$
at the bifurcation point, where lines with 
increasing dash length represent $e_{\rm bif}$ values of
$0.01,0.1,0.2,0.3,0.4,0.5,0.6,0.7,0.8,$ and $0.9$, respectively.
Corresponding colour dots represent the value of $\Psi_{\rm bif,fmin}$
for a given $e_{\rm bif}$, when just one value of $f_{\rm bif}$
satisfies Eq. (\ref{psibif}).  Adiabatic systems approaching 
the bifurcation point would be traveling upwards on this
plot while circulating nearly parallel to the X axis.
}
\label{bifurc1}
\end{figure}

The maximum value of $\Psi_{\rm bif,min}$ (obtained in the limit $e_{\rm bif} \rightarrow 1$) is $4/(3 \sqrt{3})$.  
Therefore, for any given $\Psi_{\rm bif} > 4/(3 \sqrt{3}) \approx 0.77$, there are two possible values of 
$f_{\rm bif}$ which satisfy Eq. (\ref{psibif}).  Figure \ref{bifurc1} demonstrates Eqs. (\ref{firb})-(\ref{lasb}) graphically
by plotting $\Psi_{\rm bif}$ vs. $f_{\rm bif}$ 
for 10 values of $e_{\rm bif}$
($0.01,0.1,0.2,0.3,0.4,0.5,0.6,0.7,0.8,$ and $0.9$).  Large dots mark where $\Psi_{\rm bif} = \Psi_{\rm bif,fmin}$.
As adiabatic systems increase $\Psi$ and approach the bifurcation point, their evolution can be
imagined as moving upwards on this plot while circulating almost parallel to the X-axis.  Eventually
they will reach the bifurcation point, preferentially at $\Psi_{\rm bif,min}$.
  
Note from Eq. (\ref{psibif}) that $e_{\rm bif} \rightarrow 0$ implies $\Psi_{\rm bif} \rightarrow 0$, suggesting that 
planets with initially circular orbits
can never be in the adiabatic regime.  However, this is not true.  If $e_0 = 0$, then from Eq. (\ref{adiab}), 
$e_{\rm adiabatic} = \Psi_0 \sin{f}$.  Inserting this expression into Eq. (\ref{YESmodf}) yields:

\begin{equation}
\frac{df}{dt}\big|_{t=0,e_0=0} = n\left[ \frac{\left(1 +  \frac{1}{2} \Psi_0 \sin{2f}\right)^2}{\left(1 - \Psi_{0}^2 \sin^2{f} \right)^{3/2}} - 1  \right] > 0
\end{equation}

\noindent{for} any nonzero value of $f$ (even if $f_0=0$, then $f$ attains a positive value immediately).  
After $t=0$, $df/dt$ will then continue to increase until the bifurcation point is reached. Therefore,
initially circular planets may easily evolve adiabatically, which corroborates numerical simulations.

Now we consider $d\Psi_{\rm bif}/de_{\rm bif} = 0$.  There are 3 solutions, 2 of which are physical:

\begin{eqnarray}
e_{\rm bif,ext1} &=& \frac{1}{4}  \left(  -3 \cos{f_{\rm bif}} + \sqrt{-\frac{7}{2} + \frac{9}{2} \cos{\left(2f_{\rm bif}\right)} }  \right)
\\
e_{\rm bif,ext2} &=& \frac{1}{4}  \left(  -3 \cos{f_{\rm bif}} - \sqrt{-\frac{7}{2} + \frac{9}{2} \cos{\left(2f_{\rm bif}\right)} }  \right)
\label{ecrit}
\end{eqnarray}

\noindent{where} $f_{\rm crit} \le f_{\rm bif} \le (360^{\circ} - f_{\rm crit})$, such that

\begin{equation}
f_{\rm crit} = 180^{\circ} - \frac{1}{2} \cos^{-1}\left( \frac{7}{9} \right)
\label{critf}
\approx 160.53^{\circ}
.
\end{equation}

\noindent{This} critical true anomaly value will be important for describing motion in the
runaway regime because it determines where a qualitative change in the evolution occurs.  For a given $\Psi_{\rm bif}$ and a given $f_{\rm bif}$, the number of values of $e_{\rm bif}$ which satisfy Eq. (\ref{psibif}) are:

\begin{eqnarray}
1& \ \ {\rm value} \ \ {\rm of}  \ \ e_{\rm bif} \ \ {\rm if}& \ \ 0^{\circ} \le f_{\rm bif} < f_{\rm crit}
\label{oneval}
\\
3& \ \ {\rm values} \ \ {\rm of}  \ \ e_{\rm bif} \ \ {\rm if}& \ \ f_{\rm crit} < f_{\rm bif} < 180^{\circ}
\\
\infty& \ \ {\rm values} \ \ {\rm of}  \ \ e_{\rm bif} \ \ {\rm if}& \ \ f_{\rm bif} = f_{\rm crit}
\label{infval}
\end{eqnarray}

\noindent{Limiting} values of $\Psi_{\rm bif}$ at $e_{\rm bif} = e_{\rm bif,ext1}$ and $e_{\rm bif} = e_{\rm bif,ext2}$
are $2/3$ and $4/(3\sqrt{3})$.

\begin{figure}
\centerline{
\psfig{figure=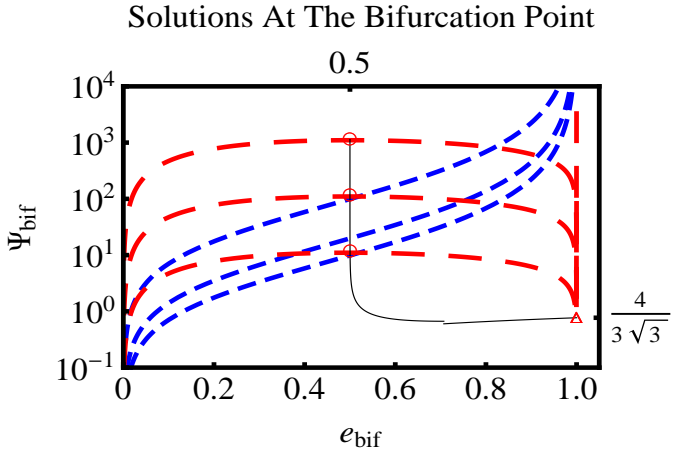,height=6cm,width=9.5cm}
}
\caption{
Values of $(\Psi_{\rm bif}, e_{\rm bif}, f_{\rm bif})$
at the bifurcation point, where the blue-short
dashed lines, starting from the top down, represent 
$f_{\rm bif} = 1^{\circ},5^{\circ},10^{\circ}$, and
the red long-dashed lines, starting from the bottom up,
represent $f_{\rm bif} = 179.0^{\circ},179.5^{\circ},179.9^{\circ}$.
The peak and trough critical points for the red long-dashed
curves are marked with red circles
and triangles, respectively.  The black curves are given
by Eqs. (\ref{crit1}) and (\ref{crit2}).  The 
blue and red curves, separated by $f_{\rm crit}$ (Eq. \ref{critf}) 
exhibit qualitatively different behaviors.  The blue short-dashed
curves and the rising portions of the red long-dashed curves 
represent evolutionary tracks {\it beyond} the 
bifurcation point, demonstrating that for 
$f_{\rm crit} < f < 360^{\circ}-f_{\rm crit}$, the planet's eccentricity
will experience an initial decrease beyond the bifurcation point.
}
\label{bifurc2}
\end{figure}

We can illustrate the multiplicity suggested by Eqs. (\ref{oneval})-(\ref{infval})
with Fig. \ref{bifurc2}.  Plotted in Fig. \ref{bifurc2} are six curves corresponding to $f = 1^{\circ}, 5^{\circ}, 10^{\circ}$ 
(short-dashed blue curves), and $f = 179.0^{\circ},179.5^{\circ},179.9^{\circ}$ (long-dashed red curves).   
We display these curves because they approximate $e-\Psi$ evolution tracks {\it beyond} the
bifurcation point in the $f \rightarrow 0^{\circ}$ and 
$f \rightarrow 180^{\circ}$ cases.
For these two values of $f$, $df/dt \approx 0$ (from Eq. \ref{YESmodf}), 
and hence
$f$ remains constant as a function of time.  In this case,
a system will move along one of these tracks.
Because $\Psi$ is always increasing, for the blue short-dashed curves, 
this system will increase its eccentricity until ejection.
However, for the red long-dashed curves, the planet's
eccentricity might {\it decrease} until reaching a 
critical point, when the increase in $\Psi$ will break
the constant $f$ approximation and force the system
off the track.  The critical points (circles; peaks) along these tracks
which appear between $1/2 < e \le 1/\sqrt{2}$ are given by:

\begin{eqnarray}
\Psi_{\rm bif,crit1} 
&=& 
\left[-\frac{6 \cos{f_{\rm bif}} + \epsilon}{24 \sqrt{3} \sin{f_{\rm bif}}}
\right] \cdot
\nonumber
\\
&&\sqrt{5 - 3 \cos{\left(2f_{\rm bif}\right)} - \epsilon\cos{f_{\rm bif}}}
,
\label{crit1}
\end{eqnarray}

\noindent{and for} those critical points (triangles; troughs) which appear 
between $1/\sqrt{2} \le e < 1$,:

\begin{eqnarray}
\Psi_{\rm bif,crit2} 
&=& 
\left[-\frac{6 \cos{f_{\rm bif}} + \epsilon}{24 \sqrt{3} \sin{f_{\rm bif}}}
\right] \cdot
\nonumber
\\
&&\frac{
\left(8-6\cos^2{f_{\rm bif}} + \epsilon\cos{f_{\rm bif}}\right)^2
}
{\left(5-3\cos{\left(2f_{\rm bif}\right)} + \epsilon\cos{f_{\rm bif}}\right)^{3/2}}
,
\label{crit2}
\end{eqnarray}

\noindent{where} $\epsilon \equiv \sqrt{18 \cos{\left(2f_{\rm bif}\right) - 14 }}$.
The two critical curves are shown on Fig. \ref{bifurc2} 
as thin black lines.  The limiting values represented by the three
triangles, which are not distinguishable by eye from one another on the plot,
occur at $e_{\rm bif} < 1$.

Linking the orbital parameters
at the bifurcation point with
the initial system orbital parameters
is difficult because although
the bifurcation point is well-defined,
the adiabatic approximations begin
to break down before the bifurcation
point is reached (see, e.g., Fig. \ref{largeevar}). 
Nevertheless,
we can analytically estimate  
the semimajor axis at the bifurcation
point by using the semimajor axis adiabat.  
Doing so gives:

\begin{eqnarray}
\frac{a_{\rm bif}}{a_0} &=& \Psi_{0}^{-\frac{1}{3}}
\left[
\frac{e_{\rm bif}^{\frac{1}{3}}\left(1+e_{\rm bif}\cos{f_{\rm bif}}\right)^{\frac{2}{3}}}{\left( 1-e_{\rm bif}^2 \right)^{\frac{1}{2}} \left(\sin{f_{\rm bif}}\right)^{\frac{1}{3}}}
\right]
\nonumber 
\\
&\propto& M_{\star}^{\frac{1}{2}}a_{0}^{-\frac{1}{2}}\alpha^{-\frac{1}{3}}
.
\label{acrit}
\end{eqnarray}

\noindent{Equation} (\ref{acrit}) contains qualitative physics
useful for understanding when the system reaches the
bifurcation point.  The dependence on the initial stellar
mass, initial semimajor axis and the mass loss rate
determine how prone a star is to 
reaching the bifurcation point and ejecting
its planet.  For two planets in separate systems
with the same $a_0$, the parent star whose 
physical parameters yield
a smaller value of $a_{\rm bif}$ is more likely
to cause ejection.  Also, wide
orbit planets are more prone to be ejected
than smaller orbit planets.  
Unfortunately, the term in square
brackets is unknown and cannot be bound without
some assumptions on $e_{\rm bif}$ and $f_{\rm bif}$.  
To be consistent with using the adiabat, one
can assume $e_{\rm bif} = e_{0}$.  However, by the time
the system has reached the bifurcation point, the 
eccentricity could have already varied away from
its initial value by at least a tenth.
The value of $f_{\rm bif}$ is the cause
of greater uncertainty.  

\subsection{``Runaway'' Regime Evolution}
\subsubsection{Runaway True Anomaly Evolution}

The unknown value of $f_{\rm bif}$ largely determines how the
planet will evolve past the bifurcation point.
If the mass loss is great and sudden enough,
then the planet will bypass the bifurcation
point altogether and immediately start evolving
in the runaway regime.  In this case, 
the planet's $f_0$ value is crucial
to its evolution.  The divided phase space structure
of Fig. \ref{bifurc2} correctly suggests that systems
can behave quantitatively differently depending on
their true anomalies.

\begin{figure*}
\centerline{
\psfig{figure=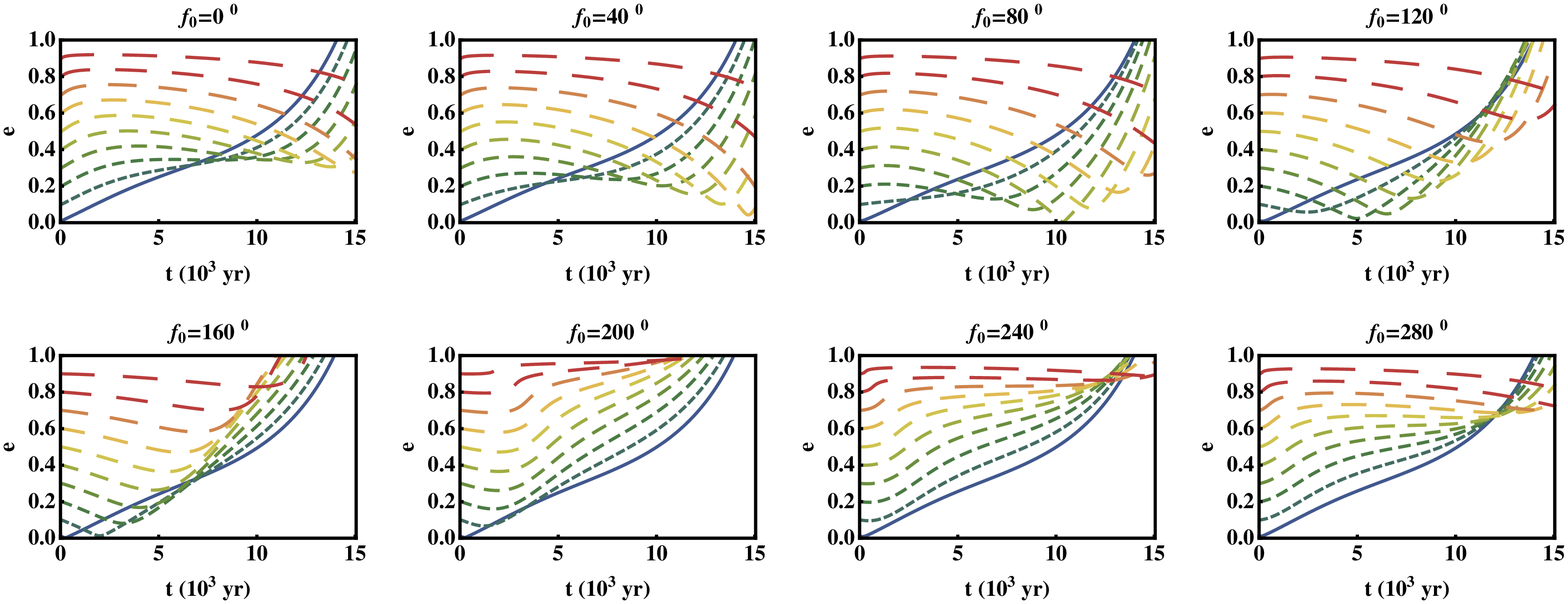,height=8cm,width=18.5cm} 
}
\caption{
How true anomaly affects eccentricity
evolution on the approach to the
runaway regime.  Shown is the eccentricity of a planet
at $a_0=500$ AU over $1.5 \times 10^4$ yr orbiting the
same star ($\mu_0 = 1 M_{\odot}$ and 
$\alpha = 5 \times 10^{-5} M_{\odot}$/yr, so $\Psi_0 \approx 0.089$) as 
in the left panels of Figs. \ref{largeevar} and \ref{avar}, as a function 
of $f_0$.  
The lines with increasing
dash length represent $e_0$ values of
$0.01,0.1,0.2,0.3,0.4,0.5,0.6,0.7,0.8,$ and $0.9$, 
respectively.
Note the dramatic sensitivity the initial true anomaly
may have on the eccentricity evolution, and 
that for $f_0=120^{\circ}-240^{\circ}$, the particle
or planet which is ejected first is one with an initial
eccentricity that is neither the highest nor lowest sampled.
The evolution is not symmetric about $f_0 = 180^{\circ}$,
and a few of the initially eccentric planets will become
circularised.
}
\label{eightplot}
\end{figure*}

\begin{figure*}
\centerline{
\psfig{figure=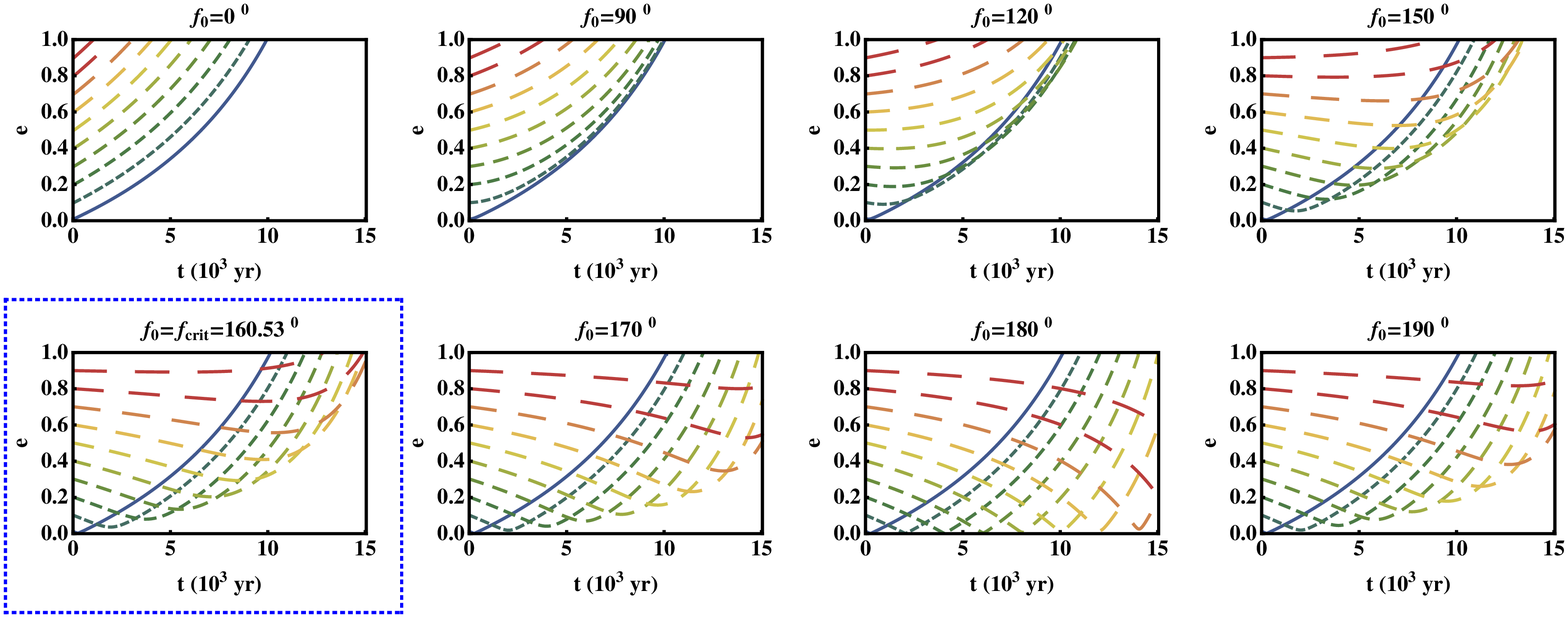,height=8cm,width=18.5cm} 
}
\caption{
How true anomaly affects eccentricity
evolution in the runaway regime.  
Shown is the eccentricity of a planet
at $a_0=10^4$ AU for the situation
in Fig. \ref{eightplot}. Here, however,
$\Psi_0 \approx 7.96$ and
the eccentricity evolution is nearly symmetric
about $f_0 = 180^{\circ}$
(as hinted at by the $f_0 = 170^{\circ}$ and $f_0 = 190^{\circ}$ cases).
Circular orbits are approached at
$f_0 = 180^{\circ}$.  The blue dashed box
highlights the case $f_0 = f_{\rm crit}$.  For 
$f_{\rm crit} < f_0 < 360^{\circ}-f_{\rm crit}$,  {\it every} planet,
regardless of $e_0$, is predicted to experience
an initial eccentricity decrease.  The eccentricity will
later increase if the mass loss continues for a long enough
time (which is not the case, e.g., for $f_0=180^{\circ}$ and $e_0=0.8$).
}
\label{fourplot}
\end{figure*}

Consider Figs. \ref{eightplot}
and \ref{fourplot}, which illustrate the eccentricity
evolution as a function of initial true anomaly
for $\Psi_0 = 0.089$ (approaching the bifurcation point)
and $\Psi = 7.96$ (runaway regime).
This dependence is more complex around the bifurcation point $\Psi \sim 0.1-1$
than after it ($\Psi > \Psi_{\rm bif}$).  Note importantly
that the divided phase space structure in Fig. \ref{bifurc2}
manifests itself strongly in Fig. \ref{fourplot} (at
$f_{\rm crit}$ in the runaway regime, highlighted by the
dotted blue box, when every planet's 
eccentricity must experience an initial decrease), but not in 
Fig. \ref{eightplot} (before the bifurcation point).
Additionally, in Fig. \ref{eightplot}, for at least a third of all 
possible initial $f_0$
values, the first planet ejected has a $e_0$ value
in between the extremes sampled of $0.01$ and $0.9$.
Contrastingly, in Fig. \ref{fourplot}, in every
instance the first planet ejected has either $e_0 = 0.01$ or $e_0 = 0.9$.
Further, the eccentricity evolution is nearly
symmetric about $f=180^{\circ}$ in the runaway regime,
a tendency not exhibited in Fig. \ref{eightplot}.  
This helps to demonstrate how complex the evolution can
be when the system is neither robustly in the adiabatic
or runaway regime.
Planets which begin their post-main sequence
life already in the strongly
runaway regime ($\Psi \gg 1$; Fig. \ref{fourplot})
experience more predictable behavior.  


Fortunately, in some cases we can analytically approximate
the evolution of orbital parameters in this regime.
If a planet begins to evolve at $t=0$ in a $\Psi \gg 1$ system
with a value of $f$ that is close to either $0^{\circ}$ or $180^{\circ}$,
then $f$ is guaranteed to librate with a small enough amplitude
so that $f$ may be treated as a constant.  Figures \ref{bifurc2}
and \ref{fourplot} suggest that the resulting behavior 
in each of the two cases
will
differ qualitatively.  The latter 
figure illustrates that
for $f = 180^{\circ}$, immediately after circularisation, the eccentricity evolution starts
increasing and continues to do so up until ejection.
Before quantifying this behavior analytically, we first attempt
to explain the physical mechanism at work:

At $f \approx 0^{\circ}$, $e$ will increase until the planet
is ejected.  There is no alternative evolutionary track.
At $f \approx 180^{\circ}$, the eccentricity will decrease
until $e \rightarrow 0$.  In this limit, $|d\varpi/dt|$
becomes large, forcing $df/dt \ne 0$.  The true anomaly
will then quickly sample other values.  At all other values
except $0^{\circ}$, $df/dt \ne 0$.  When $f$ eventually
samples $0^{\circ}$, it becomes stuck on that evolutionary
track.  


\subsubsection{Runaway Eccentricity Evolution}\label{runawayeevol}

In the runaway regime, when $f=0^{\circ}$ or $f=180^{\circ}$, Eqs. (\ref{YESmoda}) and 
(\ref{YESmode}) may be solved
directly and analytically.  
The eccentricity evolution is then given
by:

\begin{eqnarray}
e_{\rm runaway}|_{f=0^{\circ}} &=& 
e_0 \left(1 - \frac{\alpha t}{\mu_0}  \right)^{-1}
+
\left( \frac{\mu_0}{\alpha t} - 1 \right)^{-1}
\nonumber
\\
&=&
e_0 \frac{\mu_0}{\mu}
+
\left( \frac{\mu_0}{\mu} - 1 \right)
,
\label{etrueevom0}
\end{eqnarray} 

\noindent{and}

\begin{eqnarray}
e_{\rm runaway}|_{f=180^{\circ}} &=& 
e_0 \left(1 - \frac{\alpha t}{\mu_0}  \right)^{-1}
- 
\left( \frac{\mu_0}{\alpha t} - 1 \right)^{-1}
\nonumber
\\
&=& 
e_0 \frac{\mu_0}{\mu}
-
\left( \frac{\mu_0}{\mu} - 1 \right)
.
\label{etrueevom180}
\end{eqnarray}

\noindent{In} the $f_0 = 0^{\circ}$ case, the eccentricity will increase
until the planet is ejected; in the $f_0 = 180^{\circ}$ case, the eccentricity
will decrease until the planet achieves a circular orbit.
Hence, the amount of mass remaining in a star at the moment
of ejection, $\mu_{\rm out}$ and at circulation, $\mu_{\rm circ}$,
are:

\begin{equation}
\frac{\mu_{\rm out}}{\mu_0} = \frac{1 + e_0}{2}
,
\label{ejemass}
\end{equation}

\begin{equation}
\frac{\mu_{\rm circ}}{\mu_0} = 1 - e_0
.
\label{circmass}
\end{equation}

Equation (\ref{ejemass}) demonstrates that
for $\Psi > \Psi_{\rm bif}$ and $f_0 \approx 0^{\circ}$,
a planet will be ejected {\it before half of the
star's mass is lost}.  Also, planets with larger 
initial eccentricities
would be the first to be ejected.  
Equation (\ref{circmass}) demonstrates that
a planet of any eccentricity may be circularised,
and that nearly initially circular planets
are the most likely to do so first.
These equations may also be expressed as 
$t_{\rm out} = \mu_0 \left(1 - e_0\right)/(2 \alpha)$
and $t_{\rm circ} = \mu_0 e_0/\alpha$.

After circularisation, the planet's true anomaly
quickly becomes $0^{\circ}$, as described in the last subsection.
Then, the eccentricity evolves according to
a ``post-circular'' prescription:

\begin{equation}
e_{\rm post-circular} = 
\frac{t - \frac{\mu_0 e_0}{\alpha}}
{\frac{\mu_0}{\alpha} - t}
=
\frac{\mu_0}{\mu} \left(1-e_0\right)-1
.
\label{epostcirc}
\end{equation} 

\noindent{}The total time taken
for a planet to circularise {\it and then} be ejected
from a system is $t_{\rm tot} = \mu_0 \left(1 + e_0\right)/(2 \alpha)$.
After this time, the amount of material the star
has depleted is:

\begin{equation}
\frac{\mu_{\rm tot}}{\mu_0} = \frac{1 - e_0}{2}
.
\label{ejetotal}
\end{equation}

\begin{figure*}
\centerline{
\psfig{figure=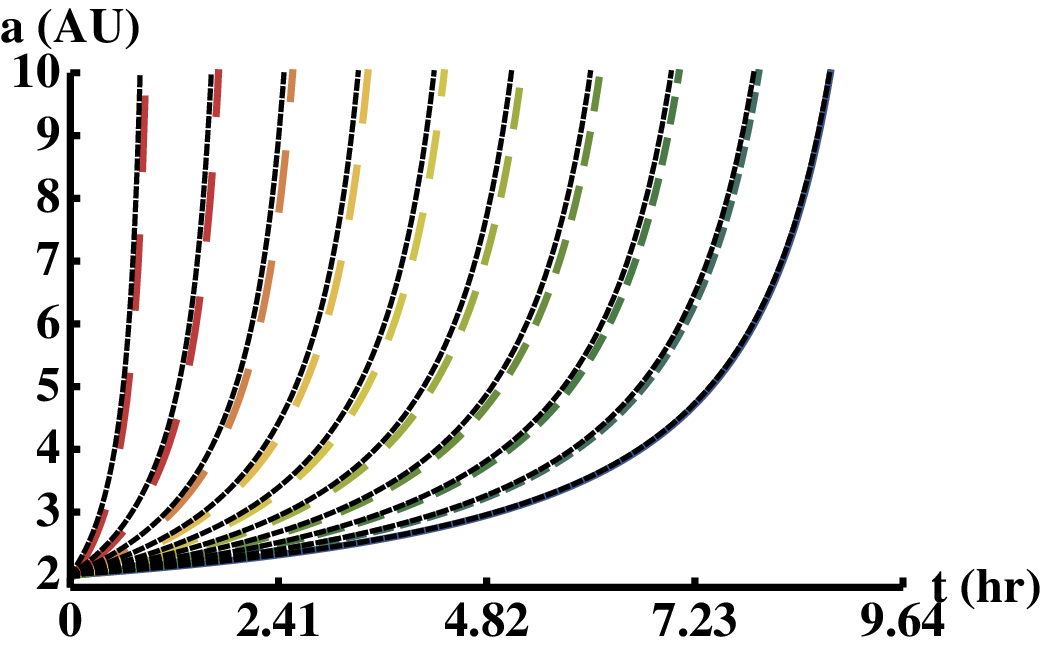,height=6cm,width=8.5cm}
\psfig{figure=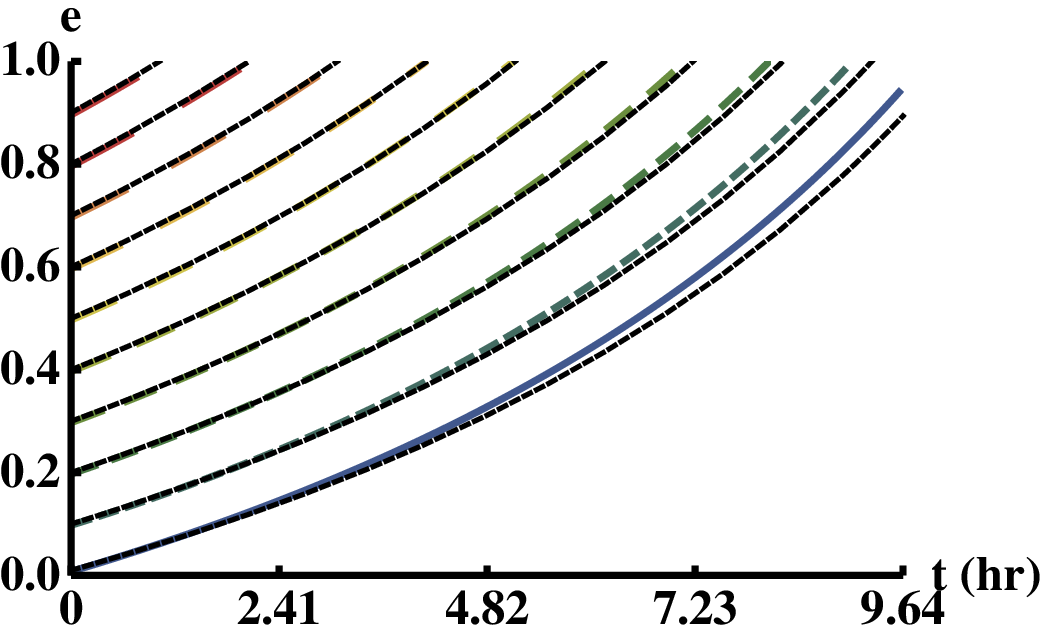,height=6cm,width=8.5cm}
}
\caption{
Analytic approximations (thick black nearly-solid foreground lines, 
from Eqs. \ref{etrueevom0} and \ref{atrueevo0}) to the $e$ and $a$ evolution
from the numerical simulations (background-coloured dashed curves)
in a robustly runaway regime for $f_0 = 20^{\circ}$.
The planet at $a_0 = 2$ AU is experiencing 
supernova-like mass loss of $\alpha = 0.5 M_{\odot}$/hour 
from a $\mu_0 = 10M_{\odot}$ star ($\Psi_0 \approx 62.4$).  The lines with increasing
dash length represent $e_0$ values of
$0.01,0.1,0.2,0.3,0.4,0.5,0.6,0.7,0.8,$ and $0.9$, 
respectively.  The analytical approximation is best for $e_0 = 0.01$,
and reproduces all the $e_0$ curves from the full numerical 
integrations to within 10\%.  All the planets
are ejected before half of the star's mass is lost (see Eq. \ref{ejemass}).
}
\label{SNf0}
\end{figure*}
\begin{figure*}
\centerline{
\psfig{figure=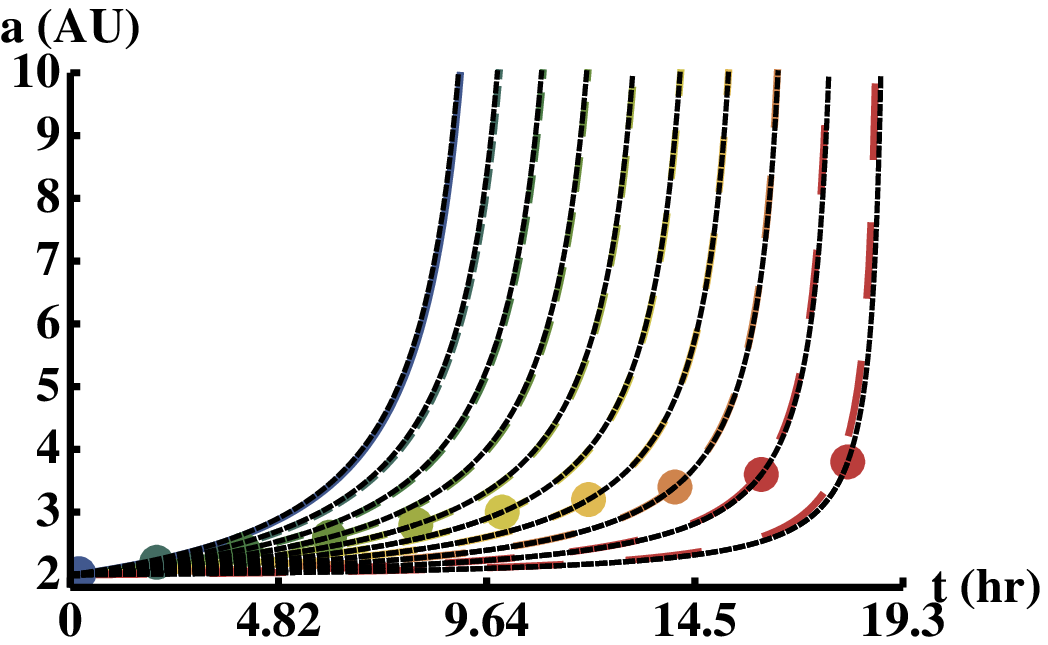,height=6cm,width=8.5cm}
\psfig{figure=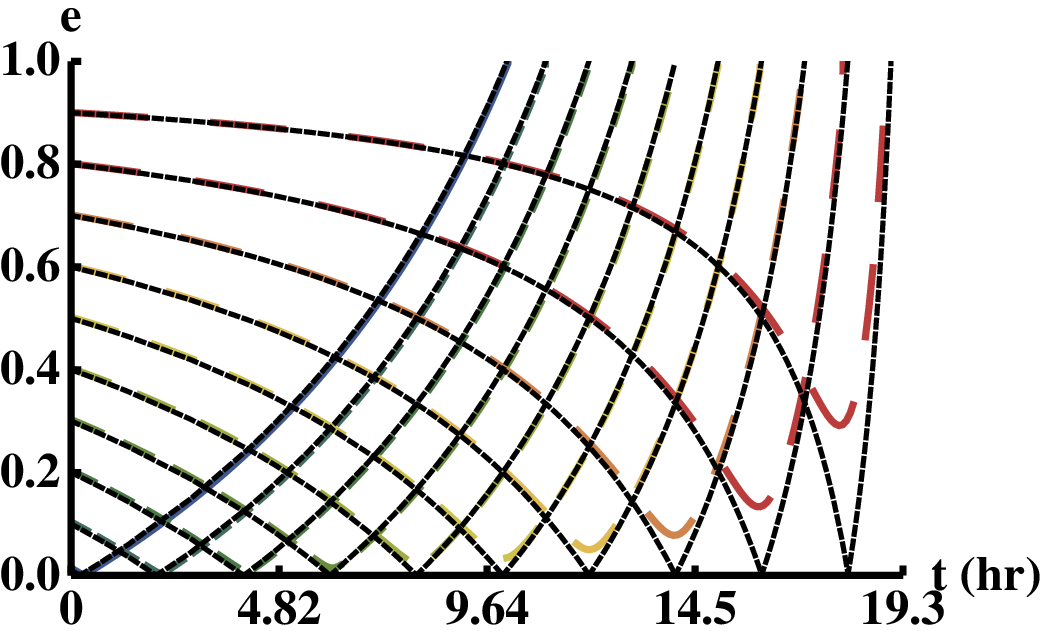,height=6cm,width=8.5cm}
}
\caption{
Analytic approximations (thin black nearly-solid foreground lines, from
Eqs. \ref{etrueevom180}, \ref{epostcirc} and \ref{atrueevo180}) 
to the $e$ and $a$ evolution from the numerical simulations 
(background-coloured dashed curves)
in the same runaway regime as in Fig. \ref{SNf0},
but for $f_0 = 178^{\circ}$.
In the limiting case of $f_0 = 180^{\circ}$,
the eccentricity decreases until reaching zero.
The dots in the left panel indicate when this 
would have occurred (at $a_0 \left[1 + e_0 \right]$).
None of the planets are ejected until at least
half of the star's mass is lost, and
the highest eccentricity planets are not ejected
until almost all of the star's mass is lost (see Eq. \ref{ejetotal}).
}
\label{SNf180}
\end{figure*}

\noindent{Any} planet which circularises before
becoming ejected therefore must have a parent
star that loses {\it at least half of its mass}.
Additionally, in order for the most eccentric planets 
to be ejected, they require the star to {\it lose all 
of its mass}. The implications are that no belt
of objects that are uniformly distributed in both 
true anomaly and eccentricity can all be ejected due to 
mass loss:  Regardless
of the value of $\Psi$, the highest eccentricity
bodies at $f \approx 180^{\circ}$ must survive.

\subsubsection{Runaway Semimajor Axis Evolution}

The semimajor axis evolution in the $f_0 = 0^{\circ},180^{\circ}$ 
runaway regime is:

\begin{eqnarray}
a_{\rm runaway}|_{f=0^{\circ}} &=& a_0 \left(1 - e_0\right) 
\frac{1 - \frac{\alpha t}{\mu_0}}{\left(1 - e_0 \right) - \frac{2 \alpha t}{\mu_0}}
\nonumber
\\
&=& 
\frac{a_0 \left( 1 - e_0\right)}{2 - \frac{\mu_0}{\mu} \left(1+e_0\right)}
\label{atrueevo0}
\end{eqnarray}

\noindent{and}

\begin{eqnarray}
a_{\rm runaway}|_{f=180^{\circ}} &=& a_0 \left(1 + e_0\right) 
\frac{1 - \frac{\alpha t}{\mu_0}}{\left(1 + e_0 \right) - \frac{2 \alpha t}{\mu_0}}
\nonumber
\\
&=&
\frac{a_0 \left( 1 + e_0\right)}{2 - \frac{\mu_0}{\mu} \left(1-e_0\right)}
\label{atrueevo180}
\end{eqnarray}

\noindent{respectively.} As one might expect, for initially circular 
orbits in the runaway regime, the semimajor axis evolution
is the same for $f_0=0^{\circ}$ and $f_0 = 180^{\circ}$.  Also,
in the circular limit, we can compare the semimajor axis
evolution with what it would have been in the adiabatic limit.
For a given $a_0$, 
$a_{\rm runaway}/a_{\rm adiabatic} = (2 \mu/\mu_0 - 1)^{-1}$,
which holds until $\mu = \mu_0/2$, the moment the planet
is ejected.

Similarly, using Eqs. (\ref{ejemass}) and (\ref{circmass}), one can show
that $a_{\rm out} = \infty$ and $a_{\rm circ} = a_0 \left(1+e_0\right)$.
Therefore, the circularisation semimajor axis is at most twice
the initial semimajor axis.  When a planet is circularised, it is done 
so only momentarily; it can only retain such an orbit if the mass loss is 
suddenly stopped at that moment.  For any planet that has
been circularised, one can show that the semimajor axis will 
subsequently evolve as:

\begin{equation}
a_{\rm post-circular} = a_{\rm runaway}|_{f=180^{\circ}}
.
\label{atrueev2}
\end{equation}

\noindent{Therefore}, the semimajor axis evolves through the $e=0$ transition
smoothly, without changing its functional form.  

We test the goodness of these analytical approximations
by considering a close-in planet (at $a_0 = 2$ AU) in the
robustly runaway regime of a supernova.  Consider a $10M_{\odot}$
progenitor which expels $\alpha = 0.5 M_{\odot}$/hour of mass
past the orbit the planet.  Thus, $\Psi \approx 62.4$.  When $f_0=0^{\circ}$
and $f_0=180^{\circ}$, Eqs. (\ref{etrueevom0}), (\ref{etrueevom180}) 
and (\ref{epostcirc})
replicate the eccentricity evolution.
Therefore, we set $f_0=20^{\circ}$ in Fig. \ref{SNf0} to 
show the extent of the deviation from the analytic 
approximation.  In the figure, the thin black dashed
lines represent the analytic approximation, which mimics
the true eccentricity evolution to within $10\%$ for
all values of $e_0$.  As predicted by Eq. (\ref{ejemass}), all planets are
ejected before the star loses half of its mass (at $10$ hrs).
In Fig. \ref{SNf180}, we set
$f_0=178^{\circ}$, just 2 degrees off an exact match,
because such a deviation from the analytics is more drastic
than for deviations of $f_0=0^{\circ}$.  As $f_0$
deviates from $180^{\circ}$, the eccentricity turns up
sooner, and becomes less circularised.  The approximation
will mimic the semimajor axis evolution until the point
at which the planet would have been circularised had $f_0=180^{\circ}$.
Note that these circularisation instances occur when $a$ is less than twice
its initial value, in conformity with $a_{\rm circ} = a_0 \left(1+e_0\right)$.
The dots in the left panel indicate when this circularisation
would have taken place, and show that the semimajor axis
evolution is unaffected.
As predicted by Eq. (\ref{ejetotal}), no planets
are ejected until at least half of the star's mass is lost,
and the highest eccentricity planets are ejected only 
in the limit of the star losing all of its mass (at $20$ hrs).

\subsection{Impulsive Regime Evolution}\label{impulevol}

One may treat the entirety of stellar mass loss under the impulse approximation,
when the mass is lost instantaneously.  This situation corresponds
to $\Psi_0 \rightarrow \infty$, an asymptotic runaway regime.  Let the subscripts
``i'' and ``f'' represent the initial and final values, $\mathcal{E}$
the (unconserved) specific energy of the system, and $r$ and $v$ 
the position and velocity of the planet.  Then

\begin{equation}
\mathcal{E}_i = \frac{1}{2} v_{i}^2 - \frac{G \mu_i}{r_i} = -\frac{G \mu_i}{2a_o}
\end{equation}

\noindent{and}

\begin{equation}
\mathcal{E}_f = \frac{1}{2} v_{f}^2 - \frac{G \mu_f}{r_f} > 0
\end{equation}

\noindent{assuming} that the planet is ejected.

Now assume $\mu_f = \beta \mu_i$, where $0 < \beta \le 1$.  
We can obtain a condition for ejection by eliminating $v_i = v_f$ 
from the equations and setting $r_i = r_f$.  Doing so gives:

\begin{equation}
\beta > \frac{1+e_{0}^2 + 2 e_0 \cos{f_0}}{2 \left(1+e_0 \cos{f_0}\right)}
.
\label{ImpPlot}
\end{equation}

\begin{figure}
\centerline{
\psfig{figure=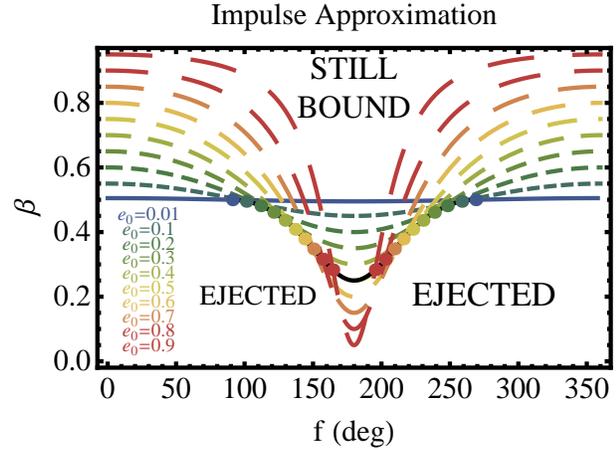,height=6cm,width=8.5cm} 
}
\caption{
Ejection prospects in the impulse approximation
($\Psi_0 \rightarrow \infty$).
Plotted is the fraction of stellar mass retained
($\equiv \beta$) vs. $f_0$ for 10 curves of increasing
dash length for
$e_0 = (0.01,0.1,0.2,0.3,0.4,0.5,0.6,0.7,0.8,0.9)$.
The planet remains bound in the regions above the curves
and is ejected in the regions below the curves.
The dots refer to the inflection points of the curves,
given by Eq. (\ref{betainf}) and which satisfy Eq. (\ref{fcrit}).
Note how highly eccentric planets are especially protected
from ejection near apocenter ($f \approx 180^{\circ}$),
but are prone to ejection near pericenter 
($f \approx 0^{\circ}$).
}
\label{ImpBlast}
\end{figure}

\noindent{We} illustrate the phase space suggested by 
Eq. (\ref{ImpPlot}) in Fig. \ref{ImpBlast}.  Below each 
curve of a given $e_0$, the planet is ejected.  Note
how the region around $f_0=180^{\circ}$ highlights
a stable region, one for which the highest eccentricity
planets are the most protected. This situation
is reflected in the finite $\Psi$ runaway regime,
and demonstrated in Figs. \ref{fourplot} and 
\ref{SNf180}.  Although the highest eccentricity
planets are the most protected at $f_0 \approx 180^{\circ}$ (apocenter),
they are the least protected at $f_0 \approx 0^{\circ}$ (pericenter).
The tendency for nearly circular planets to be ejected
is independent of true anomaly.

One curious connection between the impulse approximation and 
the bifurcation point
is that the two inflection points of Eq. (\ref{ImpPlot})
satisfy Eq. (\ref{fcrit}).  The value of
$\beta$ at these points, $\beta_{\rm infl}$, is given by:

\begin{equation}
\beta_{\rm infl} = \frac{1}{8} \left(5 - \sqrt{1+8e_{0}^2} \right)
.
\label{betainf}
\end{equation}

\noindent{Therefore}, $1/4 \le \beta_{\rm infl} \le 1/2$.
These points are marked as dots in Fig. \ref{ImpBlast},
and are connected by the analytic curve from Eqs. (\ref{fcrit}) 
and (\ref{betainf}).

Returning to Eq. (\ref{ImpPlot}), note that in the limit
of $f \rightarrow 0^{\circ}$ and $f \rightarrow 180^{\circ}$,
one recovers Eqs. (\ref{ejemass}) and (\ref{ejetotal}).
Additionally, if the inequality in Eq. (\ref{ImpPlot}) is solved for
$\cos{f}$ and then bounded by its maximum value, then ejection
is impossible if

\begin{equation}
\beta > \frac{1+e_0}{2}
.
\end{equation}

\noindent{This} condition demonstrates that
for an initially circular planet,
at least half of the star's mass must
be lost in order for there to be a possibility
of ejection.  For a highly eccentric planet,
however, just a slight mass loss might be
enough to eject it.  Whether or not a planet's 
high eccentricity serves as a protection mechanism
is dependent on its $f$ value, which relates
how close the planet is to pericenter or apocenter.

As an example, consider a circular
ring of massless particles uniformly distributed
in true anomaly at any separation from a star
of any nonzero mass.  If over half of the star's
mass is lost instantaneously, then the entire particle
ring will be ejected.  Otherwise, all the particles
will remain bound.  Now consider an eccentric ring
where all particles have $e = 0.9$.  If the parent
star instantaneously loses $60\%$ of its mass, then 
only $\approx 11\%$ of the ring will remain bound 
to the star.

\section{Excitation and Ejection in Realistic Systems}

\subsection{Overview}

We can now apply the theory developed in Section 2 to
realistic systems.  The field of stellar evolution
is extensive and touches on several areas of astrophysics.
We cannot hope to cover the entire phase space
in detail in one paper.  However, by focusing on
a single phase of stellar evolution and considering
constant mass loss in most cases,
we will attempt to provide preliminary statistics 
and order-of-magnitude analysis for the entire 
progenitor stellar mass range up to $150M_{\odot}$.  
We perform detailed nonlinear simulations only for the
$2 M_{\odot} \le \mu_{0} \lesssim 7 M_{\odot}$ regime,
whose stellar evolutionary tracks lend themselves
well to this study.

The evolution of stars is almost entirely determined
by its ZAMS (Zero-Age Main Sequence) metallicity content 
and mass \citep{wooetal2002}.  These two factors determine how mass is lost
later in life through winds.  Because this correlation
is so poorly known, mass loss prescriptions are
often treated as a third independent parameter for
tracing stellar evolution.  
To avoid detailed modelling involving integration
of the stellar evolution differential equations,
we rely heavily on the empirical algebraic fits 
of \cite{huretal2000} to model the evolutionary 
tracks of stars of most
mass, metallicity and mass loss rate properties.
These evolutionary tracks demonstrate that mass loss 
i) can occur in multiple stellar phases, ii) is 
often prominent in just one stellar phase, and iii) 
is always monotonic but typically nonlinear in any 
given phase.  All stellar evolutionary phase 
names used here are defined in their seminal work.

We use the mass loss prescriptions provided in \cite{huretal2000},
which include the Reimers law on the Red Giant Branch
\citep[RGB;][]{kudrei1978}, a steady superwind asymptotic giant branch prescription 
\citep{vaswoo1993}, a high-mass loss prescription \citep{niedej1990},
a Wolf-Rayet-like mass loss prescription \citep{hametal1995} and a luminous
blue variable law \citep{humdav1994}.  The Reimers prescription
is in particular widely used for giant branch evolution,
and beyond:

\begin{equation}
\frac{dM_{\star}}{dt} = \eta \left(4 \times10^{-13}\right)
\frac{L_{\star}(t) \mathcal{R}_{\star}(t)}{M_{\star}} \frac{M_{\odot}}{\rm yr}
\end{equation}

\noindent{where} $L_{\star}$ and $\mathcal{R}_{\star}$ are 
the stellar luminosity and radius and $\eta$ is
a dimensionless coefficient.   We adopt the 
``typical'' value for $\eta$ of $0.5$ 
\citep{huretal2000,schcun2005}.

We divide the stellar mass phase space into 5 regimes,
which are approximately separated at $1 M_{\odot}$, 
$2 M_{\odot}$, $7 M_{\odot}$, and $20 M_{\odot}$,
based on stellar evolutionary properties.

\subsection{Numerics and Checks}

Although certain regimes of evolution can
be modelled well by algebraic formulas,
the lack of a complete closed-form analytical solution to 
Eqs. (\ref{YESmoda})-(\ref{YESmodf}) suggests that 
numerical integrations are needed to model the evolution 
of the full two-body problem with mass loss.
We evolve planetary orbits in this section
by using numerical integrations.

In these integrations, one may incorporate mass loss i) as a separate differential 
equation, ii) by explicitly removing mass from the primary 
according to a given prescription, or, alternatively iii) by adding mass to 
the secondary \citep{debsig2002}.  As a check on our results, we 
have reproduced each curve in the breaking of adiabaticity regime
in the left panel of Fig. \ref{largeevar} ($\Psi_0 \approx 0.023$) with both 
i) integration of the orbital elements (Eqs. \ref{YESmoda}-\ref{YESmodf}) 
plus integration of a separate mass loss differential equation in 
the {\it Mathematica} software program, for 13 digits of accuracy and precision 
and with a working precision equal to machine precision, and ii) integration of 
the Cartesian equations of motion with the hybrid integrator of the N-body code, {\it Mercury} 
\citep{chambers1999}, with a maximum timestep of 1 yr and with mass explicitly 
being removed from the primary at each timestep. 

However, we warn future investigators that in systems which ultimately do not obey 
the adiabatic approximation, the dynamical evolution is sensitive to the evolution 
of $f$.  Therefore, in numerical integrations, particularly for nonlinear mass 
loss prescriptions, how one discretises the continuous mass-loss process can 
qualitatively affect the resulting evolution.  A discretation of mass loss is 
mimicked in reality by instantaneous bursts of primary mass lost beyond the 
orbit of the secondary.  
Therefore, a detailed study of an individual system with a given 
mass loss prescription will require a numerical integration where the time 
between discrete decreases in the primary mass should be less than the 
(largely unknown) timespan of discontinuous patterns in the mass loss modeled.  
Here, we just seek to demonstrate the instability in the general two-body mass 
loss problem and achieve representative statistics on ensembles of systems.
In the $2 M_{\odot} \le \mu_{0} \lesssim 7 M_{\odot}$ regime, which
features nonlinear mass loss, we model planets with $50$ AU $\le a_0 \le 10^5$ AU.
Therefore, we set a maximum possible timespan of 1 yr (the same value used
to reproduce Fig. \ref{largeevar}) between mass lost; in 
order to achieve mass loss on this scale, we interpolate linearly between the 
outputs from the largely nonlinear stellar evolutionary track outputs 
from \cite{huretal2000}.  We then run the simulations with {\it Mercury's} \citep{chambers1999} 
hybrid integrator.

\subsection{The Stellar Mass Spectrum}

\subsubsection{The $\mu_{0} < 1 M_{\odot}$ regime}

Sub-solar mass stars experience quiescent deaths,
some of which are theorized to last longer than
the age of the universe.  However, stellar tracks
computed from the \cite{huretal2000} code indicate 
that the most
massive members of this group ($\mu_0 > 0.7 M_{\odot}$) 
may pass through multiple stages of evolution, 
and eject up to half of their initial 
mass in the Red Giant Branch (RGB) stage.  
Low metallicity
$\mu_0 = 0.8 M_{\odot}$ and $\mu_0 = 0.9 M_{\odot}$
stars do so on the RGB
over $\sim 100-200$ Myr.  If this mass
is lost uniformly, then $\Psi_0 \approx 0.011$,
meaning that the system is likely to start
losing its adiabatic properties.  Simulations 
of {\it constant} mass loss confirm that the 
change of the eccentricity of an Oort Cloud at 
$a_0 = 10^5$ AU will vary from
$\sim 0.01$ (for particles with $e_0=0.90$)
to $\sim 0.1$ (for particles with $e_0=0.01$).
The mass loss is not strong and quick
enough to eject the particles, and objects
with semimajor axis less than $\sim 10^4$ AU (which
would yield $\Psi \le 0.00036$) are robustly
in the adiabatic regime.  This regime of the motion
might change, however, due to nonlinear modeling.
This might reveal short bursts of mass loss causing
$\Psi$ to increase sharply over the corresponding
burst timescale.

\subsubsection{The $1 M_{\odot} \le \mu_{0} < 2 M_{\odot}$ regime}

Roughly half of all known planet-hosting stars, including the Sun, 
lie in this progenitor mass regime, motivating detailed analyses
of these systems.  We defer such analyses to future studies
because of the complex multi-phasic evolutionary path
these stars are prone to follow. 

As an example, assuming $\eta=0.5$, the Sun will eventually lose 
a total of 48\% of its original mass:
24\% during the RGB, 4\% during core-He burning,
13\% during the Early Asymptotic Giant Branch (EAGB), and 7\% during
the Thermally Pulsing Asymptotic Giant Branch (TPAGB).  All these phases of mass
loss are nonlinear and occur on different timescales.  If instead
$\eta=0.3$, then the mass loss percentages will change drastically:
13\% during the RGB, 2\% during core-He burning,
4\% during the EAGB, and 28\% during
the TPAGB.  Other examples
show that slightly increasing the progenitor mass from $1.1 M_{\odot}$
to $1.2 M_{\odot}$ can have a similarly large effect on what mass is lost when.

We can, however, provide some rough estimates of planetary
evolution through
representative numerical simulations assuming constant mass
loss over one phase.  Stars in this mass regime may lose
over $60\%$ of their original mass, most of which either
in the RGB (particularly for values of $\eta \ge 0.8$)
or the TPAGB (for lower $\eta$ and
$\mu_0 > 1.3 M_{\odot}$).  The duration of RGB 
phases for these masses are $\sim 100$ Myr, and will yield only minor
eccentricity increases at $a = 10^5$ similar to those 
from sub-Solar masses.  However, the duration of
TPAGB phases in this mass regime 
is $\sim 0.1-1.0$ Myr.  Constant mass loss over this 
period of time for $\mu_0 = 1.0 M_{\odot} - 1.3 M_{\odot}$
can cause up to 20\% of an Oort Cloud 
at $10^5$ AU ($\Psi \approx 3.0$) to be ejected, and raise the eccentricity
of an initially circular planet at $10^4$ AU ($\Psi \approx 0.096$) to $\approx 0.25$.
We obtained these figures by sampling 8 evenly spaced values
of $f_0$ for each of the following 10 values of $e_0$:
$0.01, 0.1, 0.2, 0.3, 0.4, 0.5, 0.6, 0.7, 0.8, 0.9$.
This effect is pronounced with progenitor masses 
approaching $2 M_{\odot}$ and losing up to $70\%$ of their
initial mass. 

Therefore, Oort clouds are in jeopardy of partially escaping
or being moderately disrupted in systems with similar
progenitor masses to the Sun.  The comets cannot, however,
drift into the inner regions of the system (see Eq. \ref{qevol}).  
The widest-orbit planets
at $\sim 10^4$ AU may experience a moderate eccentricity
change of a few tenths, and might be ejected depending
on the nonlinear character of the mass loss.
Future multi-phasic nonlinear 
modelling will better quantify and constrain these effects.

\subsubsection{The $2 M_{\odot} \le \mu_{0} \lesssim 7 M_{\odot}$ regime}

This mass regime is well-suited for this study because 
here, $\sim 70\%-100\%$ of a star's mass loss occurs in a single
phase, the TPAGB,
regardless of the values of $\eta$, [Fe/H], or $\mu_0$.  Therefore,
by modelling the nonlinear mass loss in this one phase,
we can make definitive conclusions about this 
region of phase space.  Additionally, the duration of this
phase is short, typically under $2$ Myr, and therefore
feasible for numerical integration of planets at distances of just a few tens 
of AU.

\begin{figure*}
\centerline{
\psfig{figure=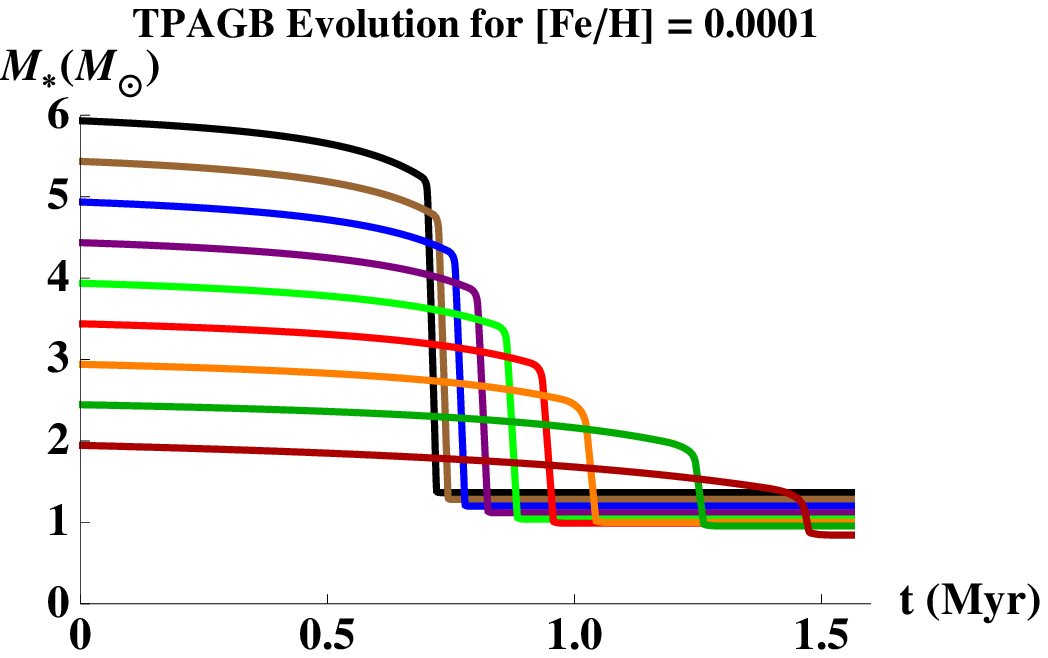,height=8cm,width=8.5cm}
\psfig{figure=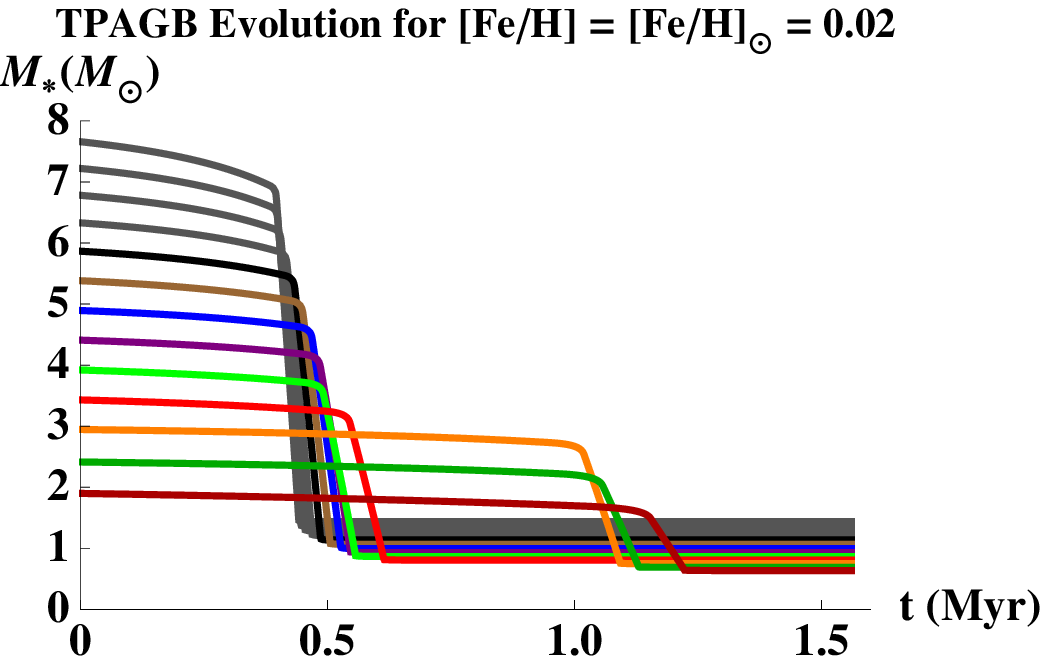,height=8cm,width=8.5cm}
}
\caption{
Thermally Pulsing Asymptotic Giant Branch (TPAGB) evolution for stars
of low- and Solar-metallicities.  Each colour represents
a different evolutionary track.  Initial TPAGB mass can be
read off from the Y-axis.  The four highest-mass
gray curves for Solar metallicities were not
computed for the low metallicity case because those
stars would have undergone supernova.
}
\label{sevo}
\end{figure*}

We consider two progenitor star metallicities, a ``low''
metallicity ([Fe/H] $= 0.0001$), and
Solar metallicity  
([Fe/H] $=$ [Fe/H]$_{\odot} = 0.02$), both with $\eta=0.5$.  In the low
metallicity case, we utilize 9 TPAGB evolutionary
tracks that range from $\mu_0 = 2 M_{\odot} - 6 M_{\odot}$, in
increments of $0.5 M_{\odot}$.
In the solar metallicity case, we utilize
13 TPAGB evolutionary tracks that range from
$\mu_0 = 2 M_{\odot} - 8 M_{\odot}$, in increments
of $0.5 M_{\odot}$.  Beyond these upper 
mass limits, a star would undergo supernova for the
stated metallicities.  The evolutionary tracks
are plotted in Fig. \ref{sevo}.  Note that
the initial masses indicated on the plots 
do not exactly represent $\mu_0$; the small ($< 10\%$) 
mass loss which occurred between
the main sequence and the start of the TPAGB, typically in the
Core-He burning and EAGB phases, has already been 
subtracted.  For all of the tracks except the low
metallicity
$\mu_0 = 2 M_{\odot}$ track,
most of the mass loss occurs within a short $10^4$ yr scale
indicated by the sharp downturn in the curves.  However,
note that between the start of the TPAGB phase to this 
intense mass loss period, over a period of $\approx 0.7-1.5$ Myr,
the stars typically 
lose $\sim 0.5 M_{\odot}$ worth of mass.  After the
intense mass loss burst, effectively no more mass is lost
from the system.  Integrations for the two lowest-mass tracks
for [Fe/H] $=$ [Fe/H]$_{\odot} = 0.02$ were begun
$5 \times 10^5$ yr after the start of the TPAGB
in order to include the sharp mass loss feature and
consistently integrate all systems over the
same period of time.

For each of the 22 evolutionary tracks, we modeled
1200 planets as test particles and integrated
the systems for $1.6$ Myr, longer than the duration of the 
TPAGB phase for nearly all of the stellar tracks.  
The planets were
all given randomly chosen values of the initial mean anomaly,
and were split into 8 groups of 150.  Each group of planets
was assigned an $a_0$ value of 
$50,100,500,1 \times 10^3,5 \times 10^3,1 \times 10^4, 5 \times 10^4,$
and $1 \times 10^5$ AU.  Each group of 150 planets was split into three subgroups 
of 50, each of which was assigned an $e_0$ value 
of $0.01$, $0.5$ and $0.9$.

\begin{figure*}
\centerline{
\psfig{figure=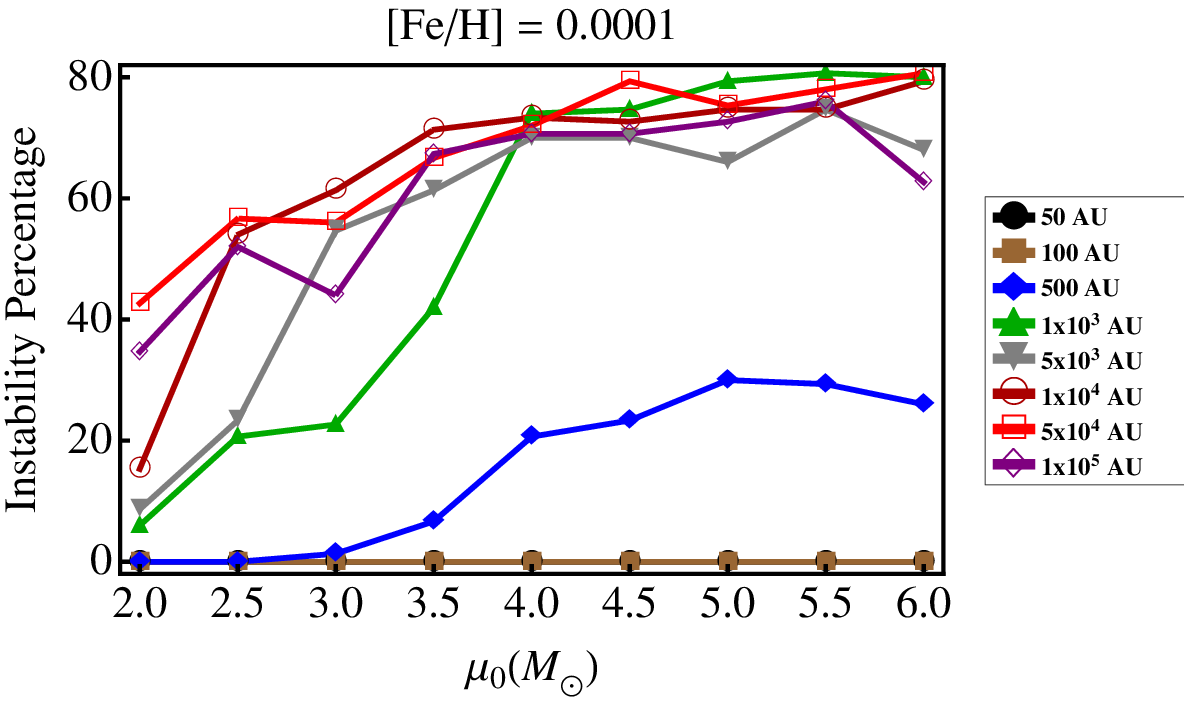,height=6cm,width=8.5cm}
\psfig{figure=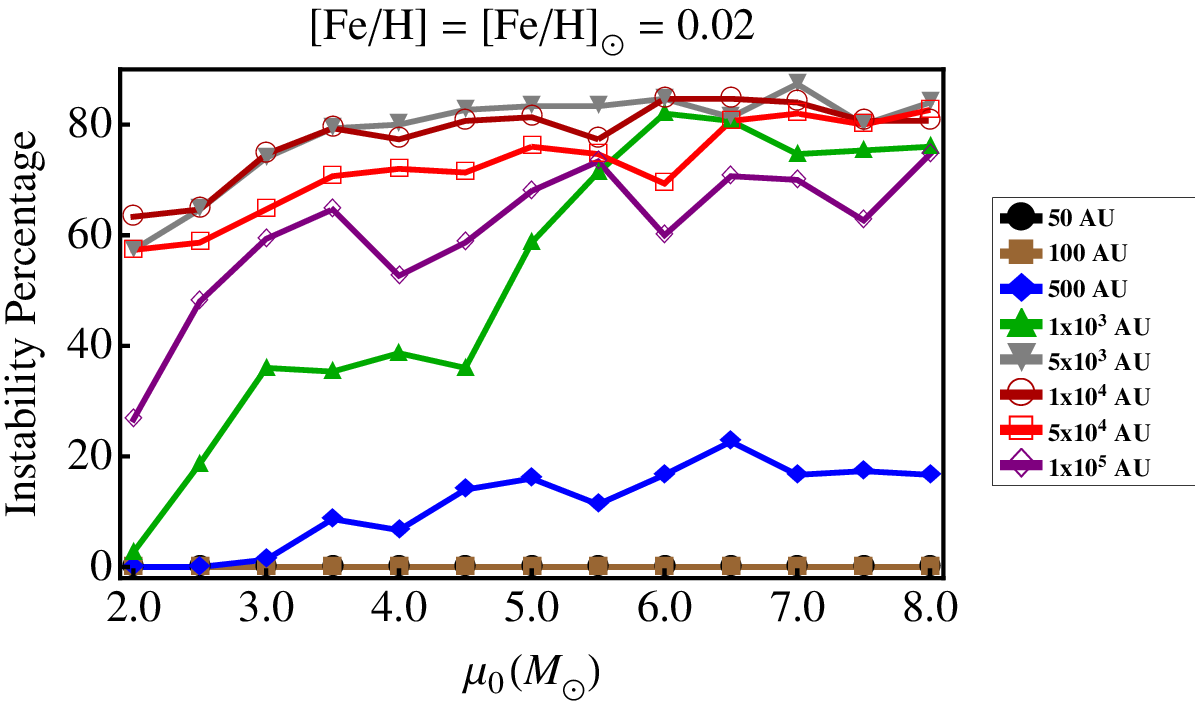,height=6cm,width=8.5cm}
}
\caption{
Planetary ejection prospects for massive stars 
from $2M_{\odot}-8 M_{\odot}$.
Each data point is averaged over the 150 randomly
chosen mean anomaly values and 3 selected $e_0$
values for each $a_0$.  The black filled-circle
curves for $50$ AU are hidden behind the $100$ AU curves. 
The $a_0 = 50$ and $a_0 = 100$ AU systems, which are in the 
adiabatic regime, remain bound.  The $a_0 \ge 10^3$ AU systems,
in the runaway regime, become largely unstable.
}
\label{Inst}
\end{figure*}

We compute the percentage of each group of 150 simulations
of a given semimajor axis and initial progenitor
mass which become unstable.  We define instability
by whether or not the planetary eccentricity reaches unity.  Figure
\ref{Inst} reports the results.  Because of the nonlinear
nature of the mass loss, here our mass loss index from Eq. (\ref{mlindex})
breaks down.  However, we can say roughly that the duration of the greatest
mass loss is comparable to a planet's period at $500$ AU
(the blue curves with diamonds).  At approximately this
semimajor axis we expect a planet to be in the transition
region between adiabaticity and runaway.  This curves
on the plot qualitatively corroborate this expectation:
orbits tighter than $500$ AU are
stable and adiabatic, orbits wider than $500$ AU
are unstable and runaway, and orbits at $500$ AU are
a bit of both.  Most of the planets in the widest orbits 
become unstable, but they cannot all become unstable
for a large enough sample of randomly chosen values of $f_0$
because some of these values will be close to $180^{\circ}$.
As demonstrated by Fig. \ref{SNf180}, in the high (e.g., 0.9) 
$e_0$ case, planets with $f_0$ close to $180^{\circ}$
will be ejected only if the parent star loses over $\sim 95\%$
of its mass, a largely unrealistic scenario for any
progenitor mass.  Further, for the simulations
in Fig. \ref{Inst}, note that beyond $1000$ AU -- in
the robustly runaway regime -- there is little correlation
with instability percentage and $a_0$.  Equations (\ref{etrueevom0}), 
(\ref{etrueevom180}) and (\ref{epostcirc}) help show why: at least 
for values of $f_0$ close to $0^{\circ}$
and $180^{\circ}$, the eccentricity evolution is {\it independent
of $a_0$}. 

\begin{figure*}
\centerline{
\psfig{figure=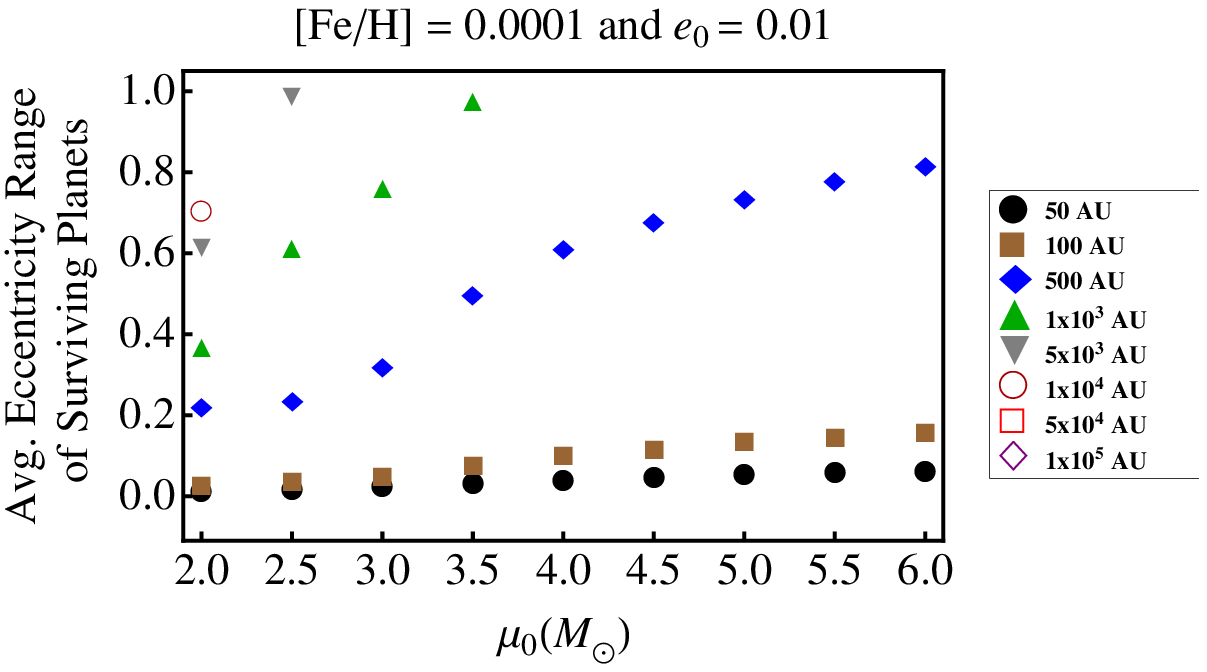,height=6cm,width=8.5cm}
\psfig{figure=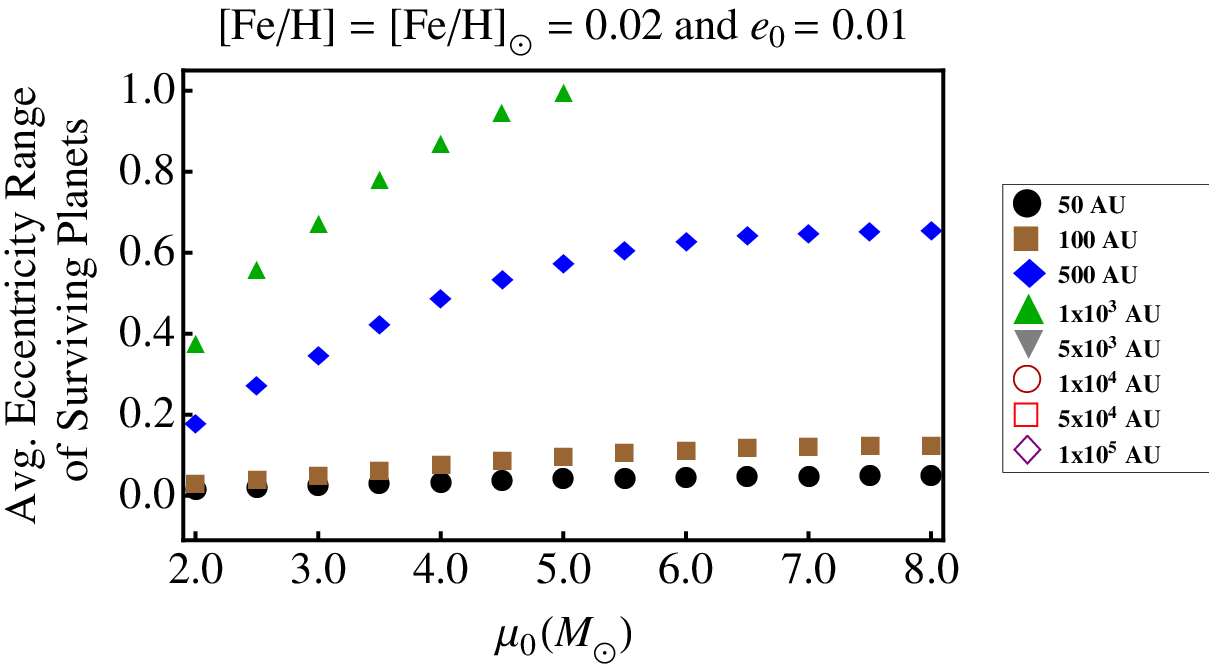,height=6cm,width=8.5cm}
}
\centerline{
\psfig{figure=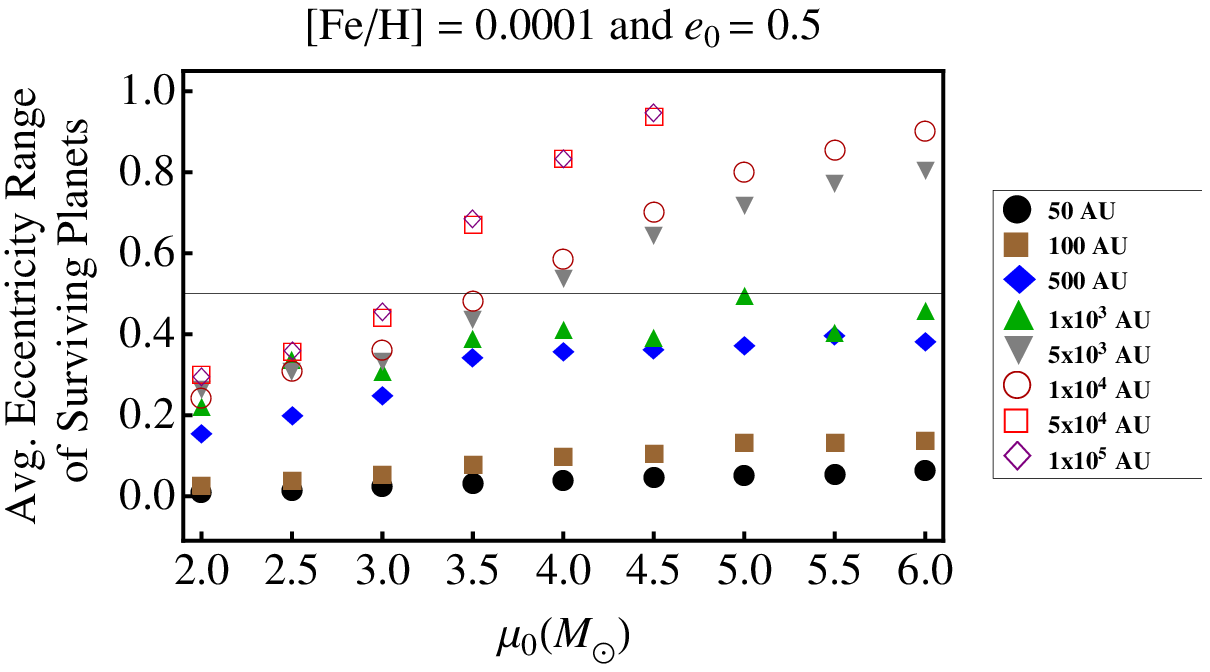,height=6cm,width=8.5cm}
\psfig{figure=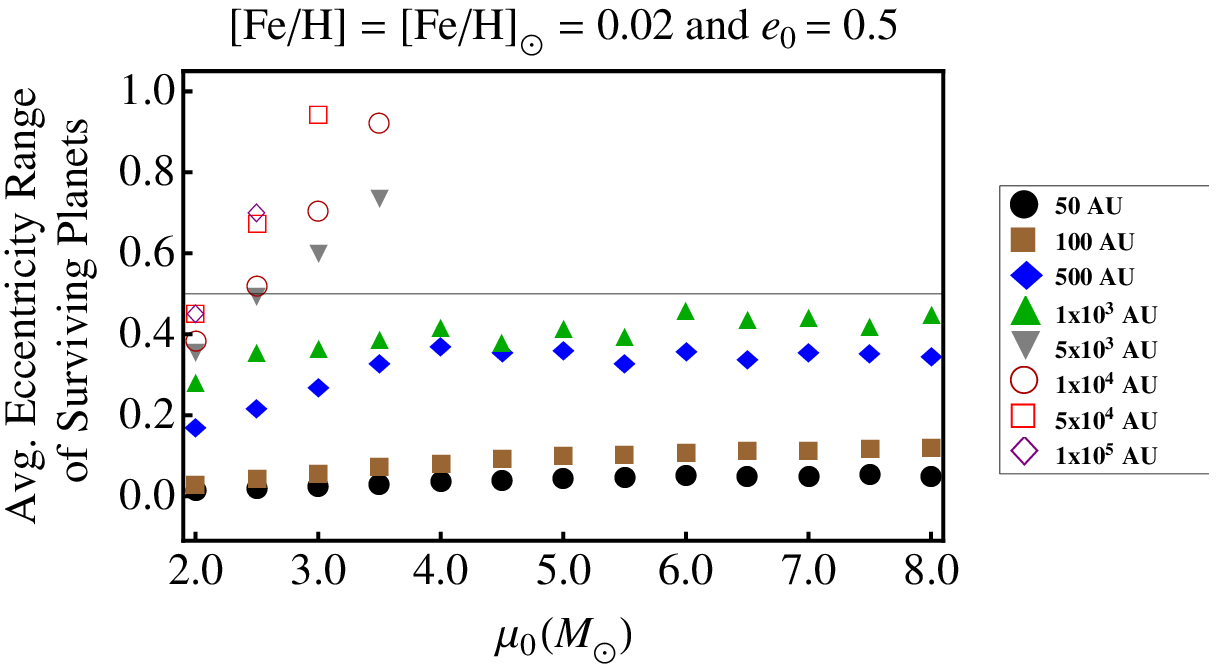,height=6cm,width=8.5cm}
}
\centerline{
\psfig{figure=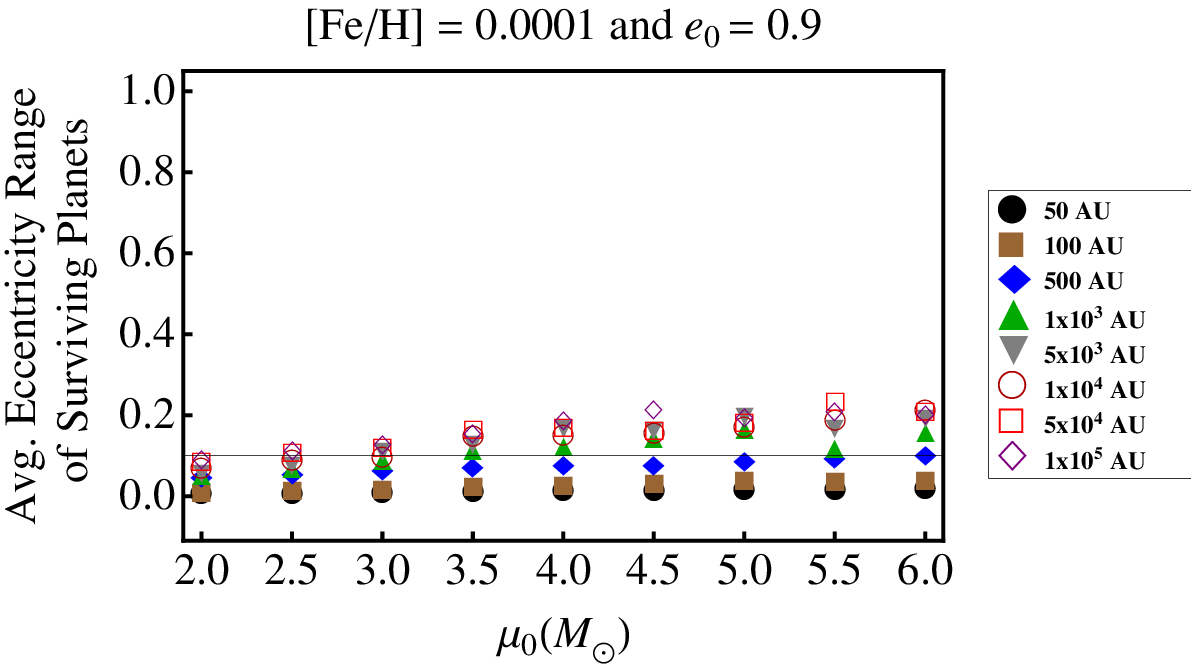,height=6cm,width=8.5cm}
\psfig{figure=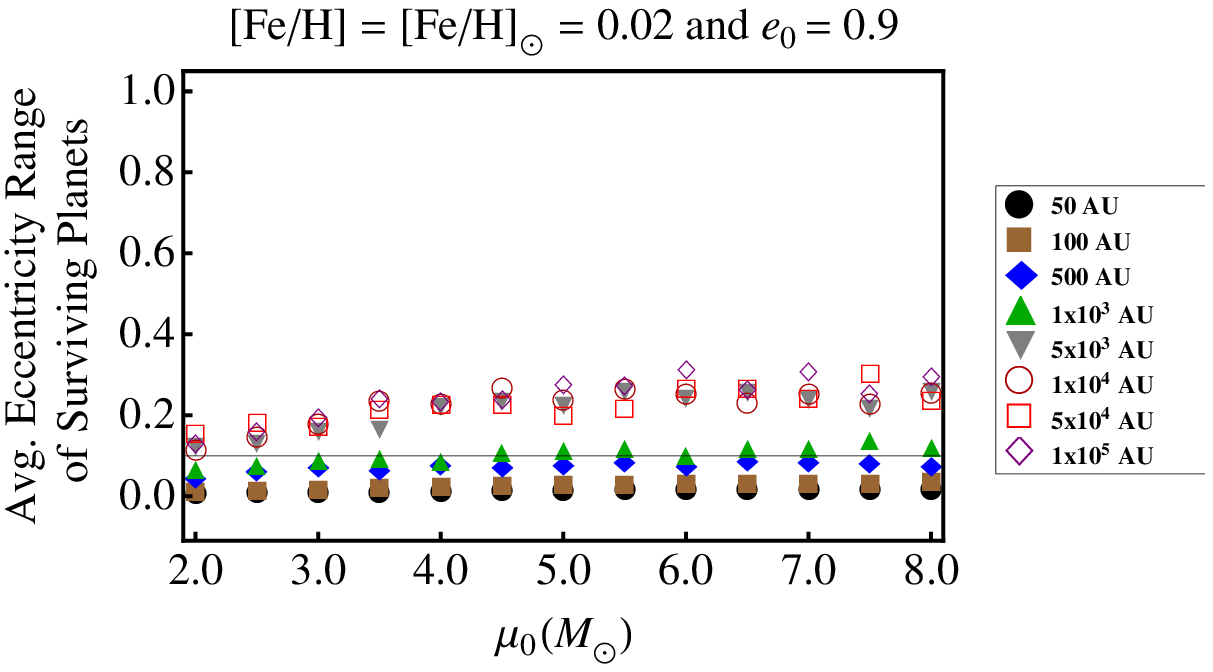,height=6cm,width=8.5cm}
}
\caption{
Eccentricity excitation of planets which remain
bound during massive star evolution, for $2M_{\odot}-8M_{\odot}$.  The left
and right panels are for low metallicity and Solar metallicity,
respectively.  The top, middle and bottom panels
are for $e_0=0.01,0.5,$ and $e=0.9$, respectively.
Each data point is averaged over the $50$ values of the mean
anomaly sampled
for the given $\mu_0$, $a_0$, and $e_0$ values.
If no symbol is displayed, then none of the corresponding
systems were stable.
The horizontal lines indicate values of $1-e_0$;
symbols above this line experience a net eccentricity
decrease.
}
\label{erange}
\end{figure*}

For the planets which remain bound, we consider
the extent of their eccentricity excitation.
Figure \ref{erange} plots the eccentricity range
experienced by bound planets averaged over
all simulations with the same values of $\mu_0$,
$a_0$, and $e_0$ but with different values of $f_0$.
The panels show that the eccentricity of the remaining
bound planets for $a_0 \ge 500$ AU is significantly
excited (by several tenths).  The eccentricity of
planets at $a_0 = 50$ and $100$ AU on average can vary by a
few hundredths, and 0.1, respectively.  If a symbol
in the legend does not appear on the corresponding plot,
then no planets at that semimajor axis remained bound.
The top two panels ($e_0 = 0.01$) exhibit a dearth of
these symbols, a result 
one might expect from Fig. \ref{ImpBlast}.  If that figure
is qualitatively representative of the situation
here, amidst strong nonlinear mass loss, then there
is no value of $f_0$ which affords the lowest
$e_0$ planets protection.
The horizontal lines on the middle two and bottom
two panels of Fig. \ref{erange} display
the value of $1-e_0$; symbols above these lines
indicate that the corresponding systems on average
experience a net eccentricity decrease.  These systems
are more likely to be left with a planet whose orbit is less
eccentric than $e_0$ when mass loss is 
terminated.

\subsubsection{The $7 M_{\odot} \lesssim \mu_{0} \lesssim 20 M_{\odot}$ regime}

Generally, Solar-metallicity stars with $8 M_{\odot} \le \mu_0 \le 20 M_{\odot}$ are
thought to undergo supernova and produce a neutron star.  However, these bounds
are approximate.  Additionally, lower metallicity stars
can begin neutron star formation and black hole formation at different
values; representative ones might be $6 M_{\odot}$ and $18 M_{\odot}$, respectively 
\citep{hegetal2003,eldtou2004,beletal2010}.
These stars may eject $\sim 50\% - 95\%$ of their initial mass,
most of which is in the supernova \citep{smaetal2009}.  Additionally, the minimum and
maximum possible masses of the remnant neutron stars are constrained by physical 
principles.  Typically accepted values for the minimum and maximum
are $\approx 1 M_{\odot}$ \citep{strwei2001} and $3 M_{\odot}$ \citep{kalbay1996};
\cite{claetal2002} presents observational evidence for the upper bound.
Therefore, this mass regime of stellar evolution is relatively well-constrained, 
and due to the nearly instantaneous mass loss, is very well suited for this study.

The sudden nature of the 
supernova, combined with the great extent of mass lost
compared to $\mu_0$, place any orbiting planet immediately in the runaway regime.
Therefore, we seek to determine what planets, if any, can {\it survive}
a supernova.  We hence choose parameters that favor survival, to see
if this situation is possible.  First, we select the minimum possible $a_0$.
Evolutionary tracks from \cite{huretal2000} indicate that the minimum
extent of the pre-supernova stellar envelope (including pre-supernova mass loss) 
is about $\sim 2$ AU, so we choose $a_0 = 2$ AU.

If the mass ejected from a supernova is considered to be isotropic, then 
this mass will collide with any orbiting bodies.  This collision is likely to destroy
smaller bodies.  Large and/or massive planets, however, may survive.  Those planets which do
survive might accrete some of the mass from the ejecta. Although doing so
will cause $a$ to decrease, this contribution, even at $2$ AU, is negligible compared to 
the $a$ increase from all the other ejecta that is being blown past the planet's orbit.  
We are concerned with the amount of time the mass takes to pass the diameter of the
planet. We can model mass loss in these systems by assuming an ejection velocity and 
a planetary diameter.

Observations help constrain the velocity of this ejecta.
Some diverse examples for different types of Supernovae include: 
i) \cite{fesetal2007} report Hubble Space Telescope observations
which indicate that the 120 yr average expansion velocity of SN1885
is $1.24 \times 10^4 \pm 1.4 \times 10^3 $ km/s, ii) \cite{mazetal2010} model spectra of SN2007gr,
and find that the inner $1 M_{\odot}$ of material is being ejected
at a velocity of $4.5 \times 10^3$ km/s, and iii) \cite{szaetal2011} find that
the maximum velocity of supernova ejecta of 2004dj during the nebular phase is 
approximately $3.25 \times 10^3$ km/s.  One theoretical investigation claims
that ejecta velocity can reach  $2\times 10^4$ km/s - $3\times 10^4$ km/s \citep{wooetal1993} ,
and another demonstrates that (surface) piston speeds of $1\times 10^4$ km/s - $2\times 10^4$ km/s 
``covers the extremes from a sudden (energy deposition over 1s) to a slow-developing explosion (energy deposition over $\sim 100$ ms)'' \citep{desetal2010}.
As exemplified by these examples as well as the compilation in Fig. 1 of \cite{hampin2002}, 
a typical range is $v = 10^3-10^4$ km/s; let us then assume the lower bound $v = 10^3$ km/s.  

Further, the highest
known exoplanet radius is less than twice Jupiter's 
radius\footnote{http://exoplanet.eu/}$^{,}$\footnote{http://exoplanets.org/}, so let
us assume this value for our planet. We can then test the
extremes of the total mass lost ($\equiv M_{eje}$) based on the 
progenitor mass and remnant mass bounds.

We assume the mass is blown past the orbit of the planet isotropically and 
at a constant value,
and we consider $36$ uniformly distributed values of $f_0$.  For each, 
we adopt $10$ values of $e_0$ ($0.01,0.1,0.2,0.3,0.4,0.5,0.6,0.7,0.8,0.9$).
We simulate these 360 systems in each of four scenarios: with
the two extreme values of $\mu_0$ ($6 M_{\odot}$ and $20 M_{\odot}$) and
two extreme values of the remnant mass ($1 M_{\odot}$ and $3 M_{\odot}$).
For all these cases, $\Psi_0 \approx 1-2$, placing these
systems in the weak runaway regime at $t=0$.
The reason why $\Psi_0$ is not higher for such a great
mass loss rate is because $a_0$ is so low (2 AU).
However, to determine the endpoint of orbital evolution, one needs
to combine an estimate of $\Psi_{0}$ with a mass loss duration time (or a
remnant mass, for constant mass loss), which is independent of $\Psi_0$.  
This is why the endstates can change drastically
for two systems even if their initial mass loss indicies are equivalent.

\begin{figure}
\centerline{
\psfig{figure=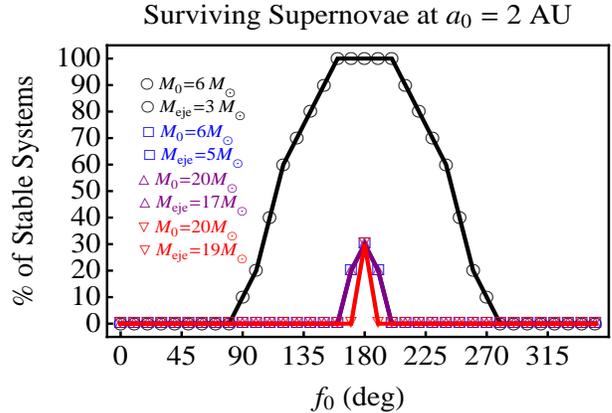,height=8cm,width=8.5cm} 
}
\caption{
Survivability of a tight-orbit ($2$ AU) planet during supernova.
The progenitor mass and ejected mass is given by 
$M_0$ and $M_{\rm eje}$, respectively.  Each data point
is based on 10 systems with 
$e_0 = (0.01,0.1,0.2,0.3,0.4,0.5,0.6,0.7,0.8,0.9)$.
For all cases, $\Psi_0 \approx 1-2$.
The plot demonstrates that the only way planets
may remain bound after a supernova blast is by initially
residing in a narrow region of true anomaly space.
This behavior was predicted in Sections \ref{runawayeevol}
and \ref{impulevol}, and specifically in  Fig. \ref{ImpBlast}.
Equation (\ref{ImpPlot}) demonstrates why {\it all}
planets on the black curve with open circles
with $160^{\circ} \le f_0 \le 200^{\circ}$ survive,
and why {\it all} planets near pericenter are ejected.
Almost all planets that withstand a supernova are ejected 
unless the percentage
of the progenitor's mass lost is the theoretical lower bound
($\approx 50\%$) for supernovae of $\approx 7 M_{\odot} - 20 M_{\odot}$ 
progenitor masses.
}
\label{SNBLAST}
\end{figure}

Figure \ref{SNBLAST} displays the result of our simulations.  The figure demonstrates
that an appreciable number of planets {\it can} survive, but only in the
extreme case of the supernova ejecting just half of the progenitor mass,
and only if $f_0$ is closer to $180^{\circ}$ than to $0^{\circ}$.  In the 
more realistic cases of greater mass loss during supernova, the only planets
which may survive must have $f_0 \approx 180^{\circ}$.  This initial
condition appears to be the only protection mechanism against ejection 
for robustly runaway (see Fig. \ref{SNf180}) or 
impulsive (see Fig. \ref{ImpBlast}) systems which lose most of 
their mass.  The impulsive limit can further help explain
Fig. \ref{SNBLAST} through Eq. (\ref{ImpPlot}): 
the top, black curve with open circles corresponds to $\beta = 1/2$,
and the highest initial eccentricity we sampled in the simulations
was $e_0=0.9$.  Therefore, Eq. (\ref{ImpPlot}) gives $\cos{f_0} < -0.9$,
meaning that {\it all} planets with $154^{\circ} \lesssim f_0 \lesssim 206^{\circ}$
should remain stable.  The numerical simulations confirm the theory.
Additionally, Fig. \ref{ImpBlast} confirms why no planets with
initial true anomalies within $80^{\circ}$ of pericenter survive,
even when just half of the star's mass is lost.

In fact, if we decrease or increase the mass loss 
rate (and hence $\Psi_0$) by an 
order of magnitude (to either
$v = 100$ km/s or $v = 10^4$ km/s), and rerun our simulations, 
we reproduce Fig. \ref{SNBLAST} closely.  Therefore, in such runaway
regimes, the evolution becomes {\it independent} of the mass loss rate
above a certain critical mass loss rate.  
The realistic implication of this finding is that
the particular choice of ejecta velocity assumed for a supernova
is unimportant, as long it is assumed to be higher than a critical
minimum value.

\subsubsection{The $\mu_{0} \gtrsim 20 M_{\odot}$ regime}

There is great uncertainty regarding how the highest-mass stars lose
mass and in what amounts.  The possibilities for planetary
evolution around these stars are intriguing, and can be simulated
once a model has been adopted for a particular star.  Stars in this 
regime generally become neutron stars or black
holes\footnote{In rare cases, at the very lowest metallicities,
pair instability supernovae will destroy the entire star
and leave no stellar remnant.}. \cite{hegetal2003} and \cite{eldtou2004} outline 
these potential stellar fates as a function of initial progenitor
mass and metallicity.

Black holes may form with or without
a supernova.  In the latter case, mass is still lost
during core collapse.  Quantifying the extent
and timescale of this mass loss is crucial for determining
the fate of any orbiting planets.  This process 
is thought to last on the order of tenths of seconds 
to seconds \citep{ocoott2011}.
The mass lost during this process has been modeled to be
as much as 1-2 $M_{\odot}$ \citep{beletal2010}.  However, this
value could be zero; \cite{fryer1999} argues that for progenitor masses
above $40 M_{\odot}$, the final black hole mass could be as large as
the progenitor. \cite{zhaetal2008} indicates
that stars at even greater masses, with $100 M_{\odot} < \mu_0 < 260 M_{\odot}$, 
may explode completely and leave no remnant.  These theoretical treatments
are poorly constrained by observations.  However,
observations do suggest that stars up to 
$300 M_{\odot}$ exist \citep{crowetal2010}.

Hence, unlike in the previous subsections, stellar mass
evolution here remains qualitatively uncertain, as the amount
of initial mass lost could be any value up to 100\%.  
Therefore, we can frame our cursory analysis in
this section by considering what {\it percent of the
progenitor's mass must be lost} in order to produce
ejection or excitation.  The large progenitor masses in this 
regime promote adiabaticity, as indicated by Eq. (\ref{mlindex}),
and hamper prospects for planetary ejection, as indicated by Eq. (\ref{acrit}).
However, mass could be lost through superwinds at a great rate
of $10^{-4} M_{\odot}$/yr \citep{desetal2010,yoocan2010}, which
might offset the stabilising effect of the large 
magnitude of the progenitor mass.

We can provide a preliminary overview of the impact
large progenitor masses with superwinds
would have on the survivability of planets.
We consider three (strong) mass loss rates, 
$\alpha = \lbrace 10^{-4}, 10^{-5}, 10^{-6} \rbrace M_{\odot}$/yr
and progenitor masses up to $150 M_{\odot}$.
For all these cases, unless $a_0 \gtrsim 10^4$ AU,
the planetary evolution will be primarily adiabatic,
as $\Psi_0 \ll 1$.  We find that at least $\sim 80\%$ 
of a progenitor's
mass must be lost for any planet at $a_0 \sim 10^3$ AU
to be ejected by any of these winds.  However,
for planets at $a_0 \sim 10^5$ AU, a mass loss
of $\alpha = 10^{-4} M_{\odot}$/yr does place the planet in a runaway
regime.  In this regime, for $f_0=0^{\circ}$, the progenitor needs to lose
just a few percent of its initial mass to eject the 
highest eccentricity planet, and roughly $50\%$ of
its mass to eject initially circular planets.
These results conform to expectation from 
Eq. (\ref{ejemass}), and hold for all progenitor
masses from $20M_{\odot} - 150M_{\odot}$.
Therefore, even
without appealing to core collapse or weak supernova,
the mass loss from the highest mass stars in the 
universe can blow away any remaining 
Oort Clouds.

Detailed modelling of secondaries
evolving amidst the complex evolution
of stars in this mass regime is a ripe 
topic for future studies.  Although the
mass lost in core collapse can approach zero,
the nearly-instantaneous timescale for the 
mass loss might have a sudden pronounced
effect on the planetary orbit.  Further,
fallback of mass from a weak supernova 
explosion onto a neutron star 
lasting ``seconds to tens of hours'' \citep{hegetal2003} 
could trigger a black hole.  This fallback
will cause a still-bound planet's semimajor 
axis to {\it decrease}.  
Also, for stars that explode away almost $100\%$
of their mass, one may investigate 
the minimum amount of mass that could
remain and still bind a planet.  
In this case, the planet's mass will
become important.

\section{Discussion}

\subsection{Oort Clouds}

No Oort Clouds have been observed.  However,
comets thought to originate in the Sun's Oort cloud have been
observed, and have motivated several studies which estimate the
orbital extent of these bodies.   \cite{levetal2010} claims the Oort cloud 
extends to $\sim 10^5$ AU, and \cite{dybczynski2002} claims that this 
is a ``typical'' value for the outer boundary.  
Although planetary material might exist throughout the scattered disk from the
Kuiper Belt to the Oort cloud \citep[e.g.][]{letetal2008}, 
some authors \citep{dunetal1987,garetal2011} have set an inner 
boundary at several thousand AU.  Other studies focus on
a supposed break in the Oort cloud, separating it into an ``inner'' and
``outer'' region.  This bifurcation is claimed to occur at
at $\sim 2 \times 10^4$ AU \citep{hills1981,kaiqui2008,braetal2010}.

These estimates pertain to the Solar System only.  Oort clouds
around other stars may exist.  Stars born in more dense clusters
will have more comets deposited into their clouds than did the Sun. 
\cite{kaiqui2008} simulate four different primordial 
environments (with no cluster, and three clusters with densities of 
$10, 30$, and $100$ stars per cubic parsec) and find that all
produce similar ``outer'' ($a > 2 \times 10^4$ AU) Oort clouds
and qualitatively different inner ones.  Further, 
\cite{braetal2010} consider the 
different types of Oort clouds which may
be formed around other stars as a function of galactocentric distance.  
At large galactocentric distances ($>14$ kpc), they find that 
some ($>10\%$) Oort cloud constituents orbit beyond $10^5$ AU.  

All these estimates suggest that
the majority of stellar mass progenitors, including the Sun and those
of sub-solar mass, will excite the eccentricity of Oort Clouds
during stellar evolution.  Most of these Oort Clouds will lose 
material to interstellar space.
Assuming that the comets are roughly distributed uniformly in true
anomaly, then only a fraction will survive.  This fraction
is highly dependent on the duration of mass loss.  The remaining
comets will assume a differential eccentricity distribution.
\cite{braetal2010} focus on galactic tides and how they strip off Oort cloud
constituents.  Indeed, Oort clouds may not even survive to the post 
main-sequence phase. If they do, the Galactic tide will be stronger 
relative to the star's gravity for any surviving Oort cloud objects, 
so the stable region that the Oort cloud can occupy will have shrunk 
at the same time that the bodies' orbits are expanding, potentially 
leading to even more ejections.  Further, as a star loses mass,
its gravitational influence within its stellar neighborhood will
shrink and be encroached by the potential wells of stellar neighbors.

However, as demonstrated by Fig. 3 of
\cite{higetal2007}, galactic tides often need Gyr of evolution
in order to cause an appreciable change of a comet's orbital
elements.  Short-lived massive stars won't often provide
galactic tides with this opportunity before stellar mass loss
becomes the dominant perturbation on the comets.

More detailed modeling of Oort clouds could enable
investigators to link mass loss from a white dwarf 
progenitor with the cometary population of the 
resulting white dwarf (see \citealt*{alcetal1986} and
\citealt*{paralc1998}).  Additionally, one should also
consider the difference in the stellar wind velocity at Oort
Cloud distances versus its escape velocity when it leaves the star.
As observed by \cite{debsig2002}, because the wind crossing time
is typically longer than the Oort Cloud orbital timescale,
winds which have slowed will enhance the system's adiabaticity.

\subsection{Wide-orbit Planets}

Initially, exoplanet discovery techniques were not well-suited
for detecting planets which reside beyond $\approx 6$ AU on decade-long
timescales, and this region
remained relatively unexplored until the mid-2000s.  However, new
observational techniques and carefully targeted surveys
are increasing the likelihood of uncovering planets on wide orbits
\citep[e.g.][]{crejoh2011}. The discoveries of the four 
planets with $a \approx 15,24,38,68$ AU orbiting
HR 8799 \citep{maretal2008,maretal2010} and the $a \approx 115$ AU planet
orbiting Formalhaut \citep{kaletal2008} revealed that wide-orbit ($a > 10-100$ AU)
planets do exist and incited great interest in their formation
and evolution.  Additionally, at least 10 wider-orbit companions which may
be massive planets that are close to the brown dwarf mass limit have been detected.
Like Formalhaut b, the companion to GQ Lup \citep{gunetal2005} is thought to 
satisfy $100$ AU $< a <$ $200$ AU.  Companions around AB Pic \citep{chaetal2005},
Oph 11 \citep{cloetal2007} and CHXR 73 \citep{luhetal2006} all harbor semimajor axis
between $200$ AU and $300$ AU, and those orbiting 
CT Cha \citep{schetal2008}, 1RXS J160929.1-210524 \citep{lafetal2010}
and GSC 06214-00210 \citep{ireetal2011} satisfy $300$ AU $< a <$ $500$ AU.
Companions with $500$ AU $< a <$ $1000$ AU include those orbiting
UScoCTIO 108 \citep{bejetal2008}, HIP 78530 \citep{lafetal2011}
and HN Peg B \citep{legetal2008}.  The three potentially planetary
companions with the widest known orbits are SR 12 C \citep[1100 AU,][]{kuzetal2011},
Ross 458 b \citep[1168 AU,][]{goletal2010}, and WD 0806-661B b \citep[2500 AU,][]{luhetal2011}.
Theoretical models place the mass of the $a = 2500$ AU object at 7 Jupiter masses.
Our study is particularly relevant to such wide-orbit companions.


During stellar evolution, wide-orbit planets in isolation will behave equivalently to
Oort cloud comets, and can be ejected with similar ease.
Planets may be mutually scattered out to distances of $\sim 10^5$ AU
while remaining bound to their parent systems \citep{veretal2009}; beyond
this distance, over time the effects of passing stars are likely
to strip the planet from the system.  Another mechanism
for producing wide-orbit planets is capture from other stars,
or passing free-floaters.  There is still a possibility
that the Sun contains a massive, very-wide orbit companion.
\cite{fernandez2011} discusses 
the prospects for detecting a wide-orbit
($>10^4$ AU) Jovian mass companion to the Sun, and \cite{matwhi2011} 
suggest that there is evidence for such a companion residing in
the Sun's outer Oort cloud.  Regardless, such planets
are very unlikely to have formed in these environments; neither
core accretion nor gravitational instability formation
models can fully form planets beyond $\sim$100 AU \citep{dodetal2009}.
Embryos and/or partially-formed planets that were scattered
beyond $\sim 10^3-10^5$ AU will undergo the same dynamical
evolution due to stellar mass loss as a fully-formed planet.
This situation might arise around short-lived, high-mass
stars, where the timescale
for core accretion might be longer than the mass loss
timescale.

\subsection{Multiple Planets}

Introducing additional bodies in the system,
such as a second planet, or a belt of material,
could significantly complicate the evolution.  \cite{debsig2002}
investigate the first scenario, and \cite{bonetal2011} the second.
In both cases, the characteristics of their N-body simulations 
demonstrated that the systems they studied were in the
adiabatic regime.  In this regime, where stellar
mass loss produces quiescent adiabatic eccentricity 
excitation on the order of $\Psi_0$ (see Eq. \ref{adiab}),
the eccentricity variation of the second planet or belt particles
can then be attributed solely to the other planet.
Additionally, the orbit of the true anomaly is only
negligibly affected by mass loss in the adiabatic limit.
Thus, the main contribution of the stellar mass loss
in their studies is through the well-defined
(Eq. \ref{dadtgen2}) increase in semimajor axis of all
objects in the system.

Including additional planets in situations where $\Psi_{\rm bif}$ is
reached and/or exceeded represents several of the numerous 
potential extensions to this work.
The frequency of planet-planet scattering and the
resulting free-floating planet population in the midst of semimajor
axis and eccentricity variations from stellar mass loss are
important issues to be addressed.  Other situations to consider
are how planets may stay locked into or be broken from secular 
and mean motion resonances, and how instability timescales are affected.

\subsection{Free-Floating Planets}

The ejection of planetary material, whether it be in the form
of partially-formed planets, fully-formed planets, or comets,
might contribute to the free-floating mass present and potentially
detectable around dead stars.  Evidence for the existence of
free-floating planets has been mounting
\citep{lucroc2000,zapetal2000,zapetal2002,bihetal2009}
and was recently highlighted by a report of 
potential detections of 10 free-floating planets 
\citep{sumetal2011}.  Also, 
the capability may exist to distinguish between free-floaters
and bound wide-orbit planets up to semimajor axes of $\approx 100$ AU \citep{han2006}.

Assuming that the same amount of planetary material was distributed
equally among stars of all progenitor masses,
then $\approx 7 M_{\odot} - 20 M_{\odot}$ progenitors are by far the 
most likely stars to produce
free-floating material\footnote{One potential
indication of the origin of supernova-produced
free-floaters is their space velocities;    
neutron star ``kicks'' cause the true space
velocities of young pulsars to reflect the (high)
speed of the supernova ejecta \citep{hobetal2005}.}, followed by stars
in the $\approx 4M_{\odot}-8M_{\odot}$ progenitor mass range 
(see Fig. \ref{Inst}).
The ability of stars with $\mu_0 \gtrsim 20 M_{\odot}$
to produce free floating material is unclear
and is largely dependent on the evolutionary models used.  
For a given
progenitor mass, metal-poor and/or metal-rich stars
may be prone to ejecting planets.  However, because metal-rich stars are 
slightly more likely to harbor planets than metal-poor stars 
\citep{setetal2010}, the metal-poor stars
which are dynamically prone to planetary excitation might not 
initially harbor planets.  

Detailed modelling of the galaxy's free-floating planet population
requires 1) an initial mass function, 2) better statistics 
for planets orbiting stars other than Sun-like hosts, 3)
knowledge of how many planets inhabit wide orbits at 
for example, $a = 10^{3-5}$ AU, and
4) better knowledge
of the ability for $\approx 1M_{\odot}-2M_{\odot}$ stellar-mass
progenitors to eject planets.  Depending on these results,
stellar evolution might be the primary source of free-floating
planets.  Alternatively, if, for example, a negligible number
of planets are shown to inhabit orbits beyond $a = 10^3$ AU, 
then the dominant source of free-floating
planets would likely lie elsewhere.


\subsection{Pulsar Planets}

Our results suggest that very few first-generation pulsar
planets exist.  Such planets would have had to reside far
enough away from the expanding progenitor envelope
to not be disrupted pre-supernova, and then survive
the supernova.  Assuming a uniform distribution of
true anomalies, only $(180^{\circ}-f_{\rm crit})/180^{\circ} \approx 11\%$ 
of planets would have a fighting chance to survive due to 
the additional time they would take to initially decrease 
their eccentricities.
Even then, their initial eccentricities would have to be high
enough, and the mass loss duration short enough, to
outlast the supernova.  These results suggest that
unless pulsars can readily form planets or capture
them from other systems, pulsar planets should be
relatively rare.  

However, if the pulsar planet survived engulfment 
from the expanding pre-supernova stellar envelope,
then its semimajor axis might be small enough to remain
bound during the supernova.  There is one planet, 
HIP 13044 \citep{setetal2010}, 
who potentially could have survived residing inside its star's
envelope \citep{beaetal2011}.  The spiral-in time of the planet
could have exceeded the short duration ($\sim 100$ yr)
of the RGB expansion and engulfment, allowing the planet to survive.
If close-in ($\lesssim 1$ AU) pulsar planets survive in a similar 
way, their final eccentricities
could be any value (see Fig. \ref{SNf180}) but their 
semimajor axes will have increased by many factors. 
The three pulsar planets orbiting PSR1257+12 \citep{wolfra1992,wolszczan1994}
all have $a < 0.5$ AU. 
If they are first-generation planets,
then $a_0 \lesssim 0.1$ would have held true for each.  At such a
small semimajor axis, their resulting dynamical evolution
during supernova would be approximately in the adiabatic-runaway
transition region ($\Psi \sim 0.1-1$).  The result is that their
pre-Supernova eccentricities (which were probably nearly zero due to tidal circularisation) 
could have been excited by a few hundredths to a few tenths, but
not by enough to have suffered ejection.
Although such values fit the observations, the system
is significantly complicated by the mutual interactions
amongst all three planets, including a resonance locking.
Instead, the observed pulsar planets may be second-generation planets \citep{perets2010},
i.e., captured (or even formed) after the supernova occurred.

\subsection{Stellar Properties}

Other questions to consider focus on the star itself.  How does
non-constant multi-phase mass loss affect the results here?
Nonisotropic and/or asymmetric mass loss may have a drastic influence  
on the resulting cometary \citep{paralc1998} and planetary \citep{namouni2005,namzho2006}
evolution.  In this case, the system no longer conserves angular momentum,
and new equations of motion must be derived.
How do short bursts, periodic or not, of ejected mass accompanying pulsating
stars affect the planetary orbit?  In this case, planetary evolution 
may even undergo several transitions between the adiabatic and runaway regimes.
The expansion and/or contraction of the stellar envelope and
the resulting tidal effects on surviving planets could also
play an important role in some cases.  Tides will compete with planetary 
ejection and possibly eccentricity excitation.
Further,
planets could be expanding their semimajor axes -- and their Hill Spheres -- 
as they are experiencing tidal effects {\it and} competing with
the expanding stellar envelope.  Some exoplanets will likely
be evaporated while others will travel through the stellar
envelope, accreting mass and being subject to a possible non-isotropic
mass distribution of the stellar envelope.

\section{Conclusion}

The variable-mass two-body problem allows for the bodies 
to become unbound or highly eccentric.  The implications
of this physical principle affect all dying stellar systems
which contain any orbiting material.  Many Oort clouds 
and wide-orbit planets will have their orbits disrupted.  
The extent of the disruption depends crucially on their 
initial semimajor axes, eccentricities,
and true anomalies, and the subtleties of
stellar evolution.  Stars with progenitor masses of
$4 M_{\odot}-8 M_{\odot}$ will readily eject objects that
are beyond a few hundred AU distant, and excite the
eccentricities of the remaining bound material at that distance. 
Supernovae which produce neutron stars eject nearly 
but not all orbiting material.  Conversely, other exotic systems, 
such as those with black holes, could have easily 
retained planets during their formation.  
Stellar mass loss might be the
dominant source of the free-floating planet population,
and orbital properties of currently observed 
disrupted planets in aged systems may be tracers of the 
evolution of their parent stars.

\section*{Acknowledgments}

We thank the referee for helpful suggestions, and Mukremin Kilic, 
Christopher A. Tout and Kimberly A. Phifer for useful discussions and references.

\bsp

\label{lastpage}

\end{document}